\magnification=1200
\hsize 5.5truein
\vsize 8.00truein
\hoffset 0.25truein
\voffset 0.25truein

   \baselineskip=20pt
%
   \baselineskip=14pt
   \hsize 6.2truein
   \vsize 8.70truein
\overfullrule=0pt


\def\lap{\hbox{{\lower -2.5pt\hbox{$<$}}\hskip -8pt\raise -2.5pt\hbox{$\sim$}}}
\def\gap{\hbox{{\lower -2.5pt\hbox{$>$}}\hskip -8pt\raise -2.5pt\hbox{$\sim$}}}

\def\Mesz{M\'esz\'aros\ }
\def\Pacz{Paczy\'nski\ }

\def\BB#1\par{
  \noindent\hangindent=20pt
  { #1 }
  \par
  \medskip
}

\def\apthree{1992, in Gamma-Ray Bursts: Huntsville, 1991, eds.
W.~S.~Paciesas \& G.~J.~Fishman (New York: AIP)}

\def\aiptwo{1994, in Gamma-Ray Bursts: Huntsville, 20-22 October, 1993,
eds. G.~J.~Fishman, J.~J.~Brainerd, \& K.~C.~Hurley (New York: AIP)}

\newcount\fignumber
\fignumber=1
\def\newfig{\global\advance\fignumber by 1}
\def\figname#1#2{\xdef#1{\the\fignumber}\newfig}

%
\newcount\eqnumber
\eqnumber=1
%
\def\neweq{{\the\eqnumber}\global\advance\eqnumber by 1}
%
\def\eqnam#1#2{\xdef#1{\the\eqnumber}}

%
%
\def\lasteq{\advance\eqnumber by -1 {\the\eqnumber}\advance
     \eqnumber by 1}

\quad\quad
\def\GSCALE{\Gamma_2}
\bigskip
\bigskip
\centerline{\bf EXPANDING RELATIVISTIC SHELLS AND }
\centerline{\bf GAMMA-RAY BURST TEMPORAL STRUCTURE}
\centerline{{\bf Edward~E.~Fenimore, Claudine~D.~Madras, and Sergei~Nayakshin}}
\centerline {{D436 Los Alamos National Laboratory, Los Alamos, NM 87545, USA}}
\centerline{e-mail: efenimore@lanl.gov}

\centerline {Received..........;  accepted............}

\centerline {{\bf ABSTRACT}}

Many models of gamma-ray bursts (GRBs) involve a shell expanding at
extreme relativistic speeds. The shell of material expands in a
photon-quiet phase for a period $t_0$ and then becomes gamma-ray active,
perhaps due to inhomogeneities in the interstellar medium or the
generation of shocks.
Based on kinematics, we
relate the envelope of the emission of the event to the characteristics
of the photon-quiet and photon-active phases.  We initially assume
local spherical symmetry wherein, on average, the same conditions
prevail over the shell's surface within angles the order
of $\Gamma^{-1}$ where $\Gamma$ is the Lorentz factor for the bulk motion.
The contribution of the curvature to the temporal structure is comparable
to the contribution from the overall expansion.  As a result,
 GRB time histories from a shell
should have an envelope
similar to ``FRED'' (fast rise, exponential decay) events where
the rise time is related to the duration
of the photon-active phase and the fall time is related to the duration of
the photon-quiet phase. This result only depends on local spherical
symmetry and, since most GRBs do not have such envelopes,
we introduce the ``shell symmetry'' problem: the observed time history
envelopes of most GRBs do not agree with that expected
 for a relativistic expanding shell.

Although FREDs have the signature of a relativistic shell,
 they may not be due to a
single shell as required by some cosmological models.
Some FREDs have precursors where the peaks are separated by more
than the expansion time required to explain the FRED shape.
Such a burst is most likely explained by a central engine, that is, the
separation of the multiple peaks occurs because the central site produced
multiple releases of energy on timescales comparable to the duration of the
event. Alternatively, there still could be local spherical
symmetry of the bulk material, but with a low ``filling factor'', that is,
only a few percent of the viewable surface (which is already very
small, $4\pi\Gamma^{-2}$) ever becomes gamma-ray active.

Long complex bursts present a myriad of problems for the models.
The duration of the event at the detector is $\sim t_0/(2\Gamma^2)$.
The long duration cannot be due
to large $t_0$ since it requires too much energy to sweep up the ISM.
Nor can it be due to small $\Gamma$ if the time variation is due to
ambient objects since the density of such objects is unreasonable
($\sim 10^{18} \Gamma^{-4}$ pc$^{-3}$ for typical parameters).
Long events must explain why they almost always violate local
spherical symmetry or why they have low filling factors.

Both precursor and long complex events are likely to be ``central engines''
which produce multiple releases of energy over $\sim 100$ s.
One promising alternative scenario is one where the 
shell becomes thicker than the
radius of curvature within $\Gamma^{-1}$. Then it acts as a parallel
slab, eliminating the problems associated with local spherical symmetry.

{\it Subject Headings:} Gamma Rays: bursts - Gamma Rays: theory
\vfill\eject

\centerline{\bf 1.~INTRODUCTION}

Although gamma-ray bursts (GRBs) have been observed and studied for nearly
twenty five years, little has been concluded about the source of the bursts.
In particular, their distance could be either cosmological
(\Pacz 1995) or galactic (Lamb 1995).
The presence of photons well above the pair production threshold
(e.g., 18 GeV, Hurley et al.~1995) has deepened the mystery;
only extreme relativistic motion will allow the escape of
such radiation.
The predicted amount of relativistic motion depends on estimates of
the distance
to and the size of the source.
The accepted method of estimating the size is to use the temporal
variations in the time histories together with causality arguments to set
an upper limit.
Originally Schmidt (1978) estimated the Lorentz $\Gamma$ factor based
on the lack of observed photon-photon attenuation and the causality argument
that the size of an object
is limited to $c\Delta T_p$ where the time $\Delta T_p$ is the duration
of a peak within a burst.   
(Here, the Lorentz $\Gamma$ is $(1-\beta^2)^{-1/2}$, where $\beta=v/c$
and $v$ is the bulk speed of radiating particles.)~~
Using $c\Delta T_p$ assumes a static location for the emitting surface.

When it became clear that GRBs could be at cosmological
distances (cf. Meegan et al.~1992), the resulting energy release
($10^{51}$ erg s$^{-1}$) implied a relativistic expanding shell.
Fenimore et al.~(1992, 1993a) estimated the Lorentz $\Gamma$ factor
based on a relativistic expanding shell with a
size of $\sim 2\Gamma^2c\Delta T_p$
rather than $c\Delta T_p$.  This size implies that the central
site acts as a ``central engine'' for the burst, ejecting material in
a fluctuating series of shells that cause the peaks
within the burst.  \Mesz \& Rees (1992) suggested there is only a
single release of energy at the central site resulting in one shell.
In these models, a single shell of expanding material becomes the
source of gamma-rays.
In this paper,
 we investigate ways
that fluctuations in the time history can be related to the size of
the emitting surface of a relativistic expanding shell.
We will base our restrictions on kinematics considerations.
This complements analyses based on hydrodynamic timescales (see, e.g.,
\Mesz \& Rees 1993, Sari \& Piran 1995).

A super-relativistic, expanding shell is a common
scenario for cosmological models since many energy
release mechanisms occur at the dynamic time scale for
compact objects (a small fraction of a second) (Piran, Shemi \& Narayan
1993; Katz 1994; \Mesz \& Rees 1993; Shemi 1994; Piran \& Shemi
1993) or over a few seconds (\Mesz \& Rees 1994;
Rees \& \Mesz 1994).
However, GRBs often display chaotic time histories that last
many seconds and have rapid variations on timescales which are
small compared to the duration of the event.
Thus, although the energy released at the central site produces
material which expands steadily outward, the photon production is not
constant.
\figname{\FIGDIVERSITY}{FIGDIVERSITY}
Figure \FIGDIVERSITY~shows the diversity
of GRB time histories observed by
the Burst And Transient Source Experiment (BATSE).
Many GRBs have somewhat simple temporal structure,
either single spikes such as in Figure \FIGDIVERSITY a or a
fast rise, exponential decay (or FRED, see Fig.~\FIGDIVERSITY b).
Other bursts show a series
of complex peaks as in Figure \FIGDIVERSITY c.  From each burst,
 one can estimate a
time scale of variation,
$\Delta T_p$, at each point $T_p$ within the burst.
For example, the complicated burst 2856 shows numerous peaks with
$\Delta T_p$ of $\sim 1$ s for about 150 s (see Fig. \FIGDIVERSITY c).
We denote the time of onset of a peak as $T_p$.
Although statistics or temporal resolution might hide the 
true number of peaks in burst 2856,
detailed fitting with a generalized pulse shape demonstrates that GRB
pulse width averages $\sim 0.6$ s,
so most of the individual peaks seen in
bursts are probably distinct entities (Norris et al.~1996).

Clearly, the rate of photon production is not directly related
to either the short time scale of the energy release ($\lap$ a few
seconds) or
the area of an expanding shell; the time histories
do not follow an envelope that scales
as $T^2_p$.
Rather, the photon production within the expanding shell must vary due
to other reasons.
For example, the expanding shell might run into 
the interstellar medium (ISM) resulting in temporal variations
(\Mesz \& Rees 1993).
Other models hypothesize that the gamma rays result from ambient
photons that are
up scattered by the relativistic particles.  The temporal
variations could be due to the variations in the ambient photon density
that might be found near an active galactic nucleus (AGN) (Epstein et al.
1993) or a
collapsed core of a globular cluster (Shemi 1994; Shaviv \& Dar 1995).
Alternatively, the growth and decay of relativistic shocks within the
shell could produce the observed rapid time variations (\Mesz \&
Rees 1994).

In this paper, we seek to explain the overall envelope of GRB
time histories.  Given the variety of models that have been suggested,
we must cover many possibilities.
The reader needs to be aware that, since this paper addresses the wide range
of suggested explanations for the time structure, different (or even
conflicting) assumptions will be made in different parts of the paper.
In \S 2 we develop the general technique for relating emission of the shell
to the observations.  In \S 3 we present five different scenarios 
(with different underlying assumptions) that
could explain some substructure within the bursts and/or the overall
duration of events.
In \S 4 we analyze particular types of GRB time histories
in terms of the scenarios discussed in \S 3.  In general, it is not
possible to explain the diversity of GRB time histories with a single
set of consistent assumptions.

\bigskip
\centerline{\bf 2.~EFFECTS OF DURATION AND THICKNESS }

\centerline{\bf ON TIME STRUCTURE}

Consider a single, relativistic, expanding shell characterized by the
Lorentz factor $\Gamma$ associated with the bulk motion.  We will
assume that all motion is radial and that the shell is spherically symmetric
in the frame of the explosion.
 In fact, the symmetry of expansion
matters only
over angles the order of a few times $\Gamma^{-1}$;
beaming prevents us from observing other regions of the shell.
Furthermore, the relatively large
Lorentz factor associated with GRBs cause important differences with
other situations involving relativistic motion such as AGNs.
Consider if the shell is not spherically symmetric but rather is a jet of
angular width $\theta_B$ (as assumed by several authors,
 e.g., Krolik \& Pier 1991).
  When the Lorentz factor is small
it is likely that an observer will see an edge of a jet rather than
be positioned directly within $\theta_B$ and sees the jet head on.
The opposite is true when the Lorentz factor is very large.  The chances
that one sees the side of the jet is the probability that the
angle of the  observer line of sight
is larger than $\theta_B$ but smaller than $\theta_B+\Gamma^{-1}$.
This probability is  about $(\theta_B \Gamma)^{-1}$.
Given the large energy release ($\sim 10^{50}$ erg s$^{-1}$) and large size
($>>10^{15}$ cm), it seems unlikely that there is a mechanism which
can routinely confine $\theta_B$ to pencil beams much smaller than
$\Gamma^{-1}$.
When the Lorentz factor is large, an observer either views the material
head on or probably will not see the source at all.  One cannot treat GRBs with
large Lorentz factors as the edges of a jet.
  Thus, we assume ``local spherical symmetry'' where
the out flowing material is essentially symmetric 
about our line of sight on angular scales of
$\sim \Gamma^{-1}$.

It is important to recognize that time variability of photon emission
can be characterized in three
different ways, leading to three different time scales.
First, one can measure time in the detector's (or ``laboratory'')
 rest frame (DRF).
We will ignore any potential motion of the explosion site
in our rest frame such as the motion due to the expansion of the universe;
these effects are
small compared to the effects due to the bulk motion.
Thus, the DRF is the same as the rest frame of the explosion.
The DRF quantities are denoted without a prime.  For example, $t$ is the
time in the DRF as measured from the initial release of energy at the
central site. Second, one can measure the ``proper'' time in
the comoving
frame (CMF) of the shell.  We denote quantities measured in the CMF
with a prime (e.g., $t'$).
The quantities in the DRF and CMF are related to each other through
a Lorentz transformation.
  The third time
scale concerns how time is measured in the DRF.
The rest frame time, $t$, is determined by clocks placed at each
point within the frame, and therefore is impossible to measure in practice.
Rather, a detector is placed at one point to observe temporal variations.
This third time we refer to as the ``arrival'' time
(denoted by capital letters, e.g., $T$).
In practice, the arrival time, the time as in 
Figure \FIGDIVERSITY, is the only time
scale which is observed.
The arrival time, $T$, is related to the DRF time as
$T = (1-\beta\cos\theta)t \approx t/(2\Gamma^2)$
(if there is a single expanding shell).
Here $\theta$ is the angle of the motion of the 
emitting region with respect to the
direction to the observer.
Thus, when one quotes that GRBs last tens of seconds and have microsecond
time variations, one is actually referring to arrival time, not
the time in the CMF or the DRF.
  The initial explosion is at time
 $T=t=0$ and forms a shell which expands
as $r = vt$.
Since, in most models the gamma-ray emitting phase does not start at the time
of the initial explosion, one does not know where to place $T = 0$.
At radius $r_0 = vt_0$
the shell begins to emit gamma rays.
These initial photons arrive at the detector at a time we denote as $T_0$.
  If $r_0$ is large,
$t_0$ cannot be neglected in
calculations.
In \S 4, we discuss how we use the results of this paper to assign when
the initial explosion might have occurred.

Note that we define all times, $T$, $t$, and $t'$, to be measured from
the initial energy release, so that a peak arriving, say, $30$ seconds
after the first gamma-rays, occurs at time $T_p = T_0 +30$ s.

\centerline{\it 2.1 Timescales from Thickness and Duration}

An observer viewing a spherically symmetric,
relativistically expanding shell from the vantage point of a single
position ``sees'' a shape defined by the photons that arrive at the
observer at the same time.
These photons originate from
a prolate ellipsoid (Rees 1966).
After arrival time $T$, the semimajor axis is $\Gamma^2 vT$,
the semiminor
axis is $\Gamma vT$, and the eccentricity is $\beta$.
In the rest frame of the detector, the distance from the center
of the explosion to the point closest to the detector is
\figname{\FIGELLP}{FIGELLP}
$v(1+\beta)\Gamma^2 T \approx 2\Gamma^2c T$ (see Fig. \FIGELLP).
The curvature within $\Gamma^{-1}$ is small and previous papers have neglected
it.  However, we will show that the curvature has temporal effects which
are comparable to the temporal effects of the overall expansion.  To
understand this, one must distinguish temporal effects associated with
the duration of the emission in the DRF from those effects associated
with the thickness of the emitting region in the DRM.

Consider a shell with thickness $\Delta r_{\parallel}$ (as measured
in the DRF) that emits for a duration $\Delta t$ from
\figname{\FIGTHICKDUR}{FIGTHICKDUR}
time $t_1$ to $t_2$ (see Fig.~\FIGTHICKDUR).
At time $t_1$, photon ``A'' is emitted from the leading edge of
a shell with finite thickness and photon ``B'' is emitted from the
trailing edge.
At time $t_2$, photons ``a'' and ``b'' are emitted from the leading
and trailing edge, respectively.
Since the edges are moving at speed $v$ (very close to $c$),
the photons emitted from the same edge of the shell but at
different times (i.e., ``A'' and ``a'') arrive at the detector separated
in time by $(1-\beta)(t_2-t_1) \approx (2\Gamma^2)^{-1}\Delta t =
(2\Gamma)^{-1} \Delta t'$.
Thus, the emission from single shells appears to be 
compressed in arrival time to be
$2\Gamma^2$ times shorter than in the DRF, and $2\Gamma$ times shorter
than in the CMF.

In contrast, a thick shell emits photons  simultaneously (in the DRF)
 from opposite edges,
(e.g., photons ``A'' and ``B'' in
Fig. \FIGTHICKDUR) which arrive at the detector separated 
in time by $\Delta r_{\parallel}/c$.
Thus, the emitting region of a GRB with $\sim 1$ sec peaks 
must not have been thicker than 1 light second
in the DRF.  One can conclude that at no time throughout
either burst in
Figure \FIGDIVERSITY a or 1c was the thickness of the 
emitting region greater than
approximately a
light second.  We draw this conclusion generally for GRBs; at no time
can the regions emitting photons have a
line-of-sight width greater than the observed width of the peaks.
Since the expanding shell is $\sim 2\Gamma^2 T_0$ in
size, the width of the peaks requires the emitting region to be very thin:
$\Delta r_{\parallel}/r_0 \sim \Delta T_0 (2\Gamma^2 T_0)^{-1}$
which is about $10^{-2}\Gamma^{-2}$
when $T_0$ is as large as hundreds of seconds.

Duration of emission, therefore, causes a peak to appear to be $
(2 \Gamma^2)^{-1}$ times shorter in arrival time as it is in the DRF,
while thickness of the emitting region causes a peak that appears to
the detector as it was emitted in the rest frame.  For this reason,
even a small thickness
in the emitting region can have a great effect on the observed time
histories. The delay due to the curvature is
 a thickness effect ($d_c$ in Fig. \FIGTHICKDUR).
The curvature is $\sim \Gamma^{-2}$ times smaller than the radius but
thickness effects are not compressed in time like duration effects are.
The expansion is a duration effect and is compressed by $2\Gamma^{-2}$.
Thus, the curvature has an impact comparable to the expansion.

\bigskip
\centerline{\it 2.2 Special Case: An Infinitely Thin Shell}

  The simplest scenario which
demonstrates the effect of the curvature is an
infinitely thin shell which expands in a photon-quiet stage and then emits
photons for some time $t_b$ in the DRF
where $t_b$ is the duration of the gamma-active phase.
It is clear from Figure \FIGDIVERSITY~that the 
shell of material does not emit constantly
especially for complex bursts.
To describe this analytically, let
us define a photon production rate on the
shell which varies as position and time.
Let $P(\theta,\phi,t)\,dt$ be the number fluence (photons cm$^{-2}$)
emitted
at location $\theta,\phi$ between time $t$ and $t+\,dt$.
The general expression
for the flux (photon cm$^{-2}$ s$^{-1}$) in the bandpass $E_1$ to $E_2$
observed between $T$ and $T+\,dT$ is
\eqnam{\SCV}{SCV}
$$
V(T) = {1 \over D^2}   \oint\displaylimits_{\rm{ellip}}
\int_{E_1 \Lambda}^{E_2 \Lambda}{\Phi(E{^{\prime}})  \over \Lambda}
P(\theta,\phi,t)  \Lambda^{-2}
\,dE{^{\prime}} \,dA
\eqno(\neweq)
$$
where
$$
\Lambda = \Gamma(1-\beta \cos\theta),~~~
\eqno(\neweq)
$$
$D$ is the distance to the observer, and the surface differential,
$\,dA$, lies on the surface of the prolate ellipsoid that produces
photons which arrive at time $T = t/(2\Gamma^2)$.
The spectrum of the emission in the comoving frame of the shell is
$\Phi(E{^{\prime}})$.
The $\Lambda^{-2}$ term accounts for the beaming that occurs when one
transforms from the rest frame of the expanding shell to the rest frame
of the detector.
Let $\Lambda_0 = \Gamma(1-\beta)$, then
the factor $\Lambda/\Lambda_0$ gives the relative boost of the rest frame
spectrum between those photons that arrive first (i.e., $\Lambda_0$) and at
time $T$ (i.e., $\Lambda$).
We assume that the emission in the comoving frame of the shell is isotropic.

We first try to make the most narrow time structure possible by assuming
an infinitely thin shell
becomes uniformly photon-active for
an infinitesimal time $\,dt$, i.e. $P(\theta, \phi, t) = P_0\delta (t-t_0)$.
Previously, we assumed local spherical symmetry for the out flowing
material. 
 Here, by assuming
$P(\theta,\phi,t)$ is independent of $(\theta,\phi)$, we are adding
the additional assumption of local spherical symmetry 
for the conversion of bulk energy into
gamma-rays.  Later, we will relax this assumption.
The distance from the initial explosion to a point on the ellipsoid
defined by the angle $\theta$,
 $r_{el}$, is
 $r_{el} = vt (1- \beta \cos \theta)^{-1}$.
The surface
which corresponds to the $\delta$-function emission is
characterized in arrival time as
\eqnam{\TTREL}{TTREL}
$$
r_0 = {vT_0 \over 1-\beta}~={vT \over 1-\beta\cos\theta}~.
\eqno(\neweq)
$$
Such that
\def\costerm{{1-{T\over T_0}(1-\beta) \over \beta}}
$$
\cos\theta=\costerm
\eqno(\neweq)
$$
and

$$
\Lambda = {T/T_0 \over (1+\beta)\Gamma}~~.
\eqno(\neweq)
$$
The $\,dA$ in equation (\SCV) is the same as the
$\,dA$ of a spherical shell facing the observer,
that is,
\eqnam{\SSDA}{SSDA}
$$
\,dA = 2 \pi r_0^2 \sin \theta \cos \theta \,d\theta ~.
\eqno(\neweq)
$$
Using equation (\TTREL),
$$
\,dA = {\pi r_0^2 \over \Gamma^2 T_0}\,dT = 4\pi \Gamma^2 c^2 T_0 \,dT
\eqno(\neweq)
$$
The resulting time variation for a single shell emitting for a short
time period, as seen by the observer, is
\eqnam{\VSINGLE}{VSINGLE}
\def\Vdelta{V_{\delta}(T)}
$$
\Vdelta \,dT =  0~~~~~~~~~~~~~~~~~~~~~~~~~~~~~~~~~~~~~~~~~~{\rm if}~T<T_0
\eqno(\neweq a)
$$
$$
~~~~~~~~~~~~~~~~~~~~~~~= \psi \Gamma^{4+\alpha} P_0~
\bigg({T \over T_0}\bigg)^{-\alpha-2} ~~T_0\,dT
{}~~~~~~{\rm if}~T>T_0
\eqno(\lasteq b)
$$
where we have assumed that the rest frame photon number
spectrum is a power law with index $-\alpha$ and
$\psi$ is the constant
$$
\psi = {2^{4+\alpha}c^2\pi  \over D^2}\int_{E_1}^{E_2}E^{-\alpha}dE~~.
\eqno(\neweq)
$$
Since the emission is assumed to occur at a single radius, changes in
$\Gamma$ (such as slowing down in the ISM) do not affect the shape.

\figname{\FIGVDELTA}{FIGVDELTA}
As long as $\theta \sim \Gamma^{-1}$ is small,
the time variation in $\Vdelta$ depends only on ${T/
T_0}$; the shape is shown in Figure \FIGVDELTA a for $\alpha = 1.5$.
Note that the shape depends on where $T$ is defined to be zero which
is when the central explosion occurred.
The full width half maximum (FWHM)
of $\Vdelta$
varies from $0.26 T_0$ to 0.19$T_0$ as $\alpha$ varies from 1 to 2.
Thus, the shape expected for a single shell that turns on is rather
universal, it has a weak dependency on the spectral shape.
For the purposes of examples in this paper we will use $\alpha = 1.5$
such that the FWHM of $\Vdelta \sim 0.22T_0$.
  This shape is similar to the time
histories of the GRBs referred to as FREDs, which are characterized by
their fast rise times and long tails.
We propose that FREDs result from an expanding shell that emits for a small
range of times at $t_0$.  Although called ``FREDs,'' according to
equation (\VSINGLE) the tail is not exponential but $T^{-\alpha-2}$.
If all bursts were FRED-like, then relativistic shells that turn
on after a time $T_0 = t_0(2\Gamma^2)^{-1}$ would easily
explain the envelope of emission.

\centerline{\it 2.3 Infinitely Thin Shells Emitting for Finite Time}

The shape $\Vdelta$ was derived assuming that $P(\theta,\phi,t)$ was
a $\delta$-function.  More complex envelopes can be found as weighted
sums of $\Vdelta$.  For example,
as the bulk material converts its energy to gamma rays, $\Gamma$ should
decrease.
We assume $\Gamma(T)$ can be approximated as $\Gamma_0(T_e/T_0)^{-\zeta}$
where $T_e$ is the time of emission.
If $P(\theta,\phi,t)$ has local spherical symmetry ($=P(T)$) and
is constant ($=P_0$) from
$t=T_{0} 2\Gamma^2$ to $t=T_{\rm max}2\Gamma^2$ and zero otherwise, then
$$
V(T)
 = \psi \int_{T_{0}}^{T} \Gamma^{4+\alpha}
(T_e)\bigg({T \over T_e}\bigg)^{-\alpha-2} P(T_e) T_e \,dT_e
\eqno(\neweq)
$$
such that
\eqnam{\VTHICK}{VTHICK}
$$
V(T) = 0  ~~~~~~~~~~~~~~~~~~~~~~~~~~~~~~~~~~~~~~~~~~~~~~~~~~{\rm if}~~T < T_{0}
\eqno(\neweq a)
$$
$$
{}~~~~~ =
{\psi\Gamma_0^{4+\alpha} P_0 \over \omega} 
{T^{\omega} - T_{0}^{\omega} \over T_0^{-4\zeta}T^{\alpha+2}}
{}~~~~~~~~~~~~~~~~{\rm if}~T_{0} < T < T_{\rm max}
\eqno(\lasteq b)
$$
$$
{}~~~~ =
{\psi\Gamma_0^{4+\alpha} P_0 \over \omega} 
{T_{\rm max}^{\omega} - T_{0}^{\omega} \over T_0^{-4\zeta} T^{\alpha+2}}
{}~~~~~~~~~~~~~~~~~{\rm if}~T > T_{\rm max}
\eqno(\lasteq c)
$$
where
\eqnam{\POWERINDEX}{POWERINDEX}
$$
\omega=\alpha-(4+\alpha)\zeta+4     \eqno(\neweq)
$$
If $\Gamma$ decreases to $\Gamma_{\rm min}$, then
\eqnam{\ZETA}{ZETA}
$$
\zeta \sim {\rm ln}(\Gamma_0/\Gamma_{\rm min})/{\rm ln}(T_{\rm max}/T_{0})~~.
\eqno(\neweq)
$$
where $\Gamma_{\rm min}$ is the $\Gamma$ at $T_{\rm max}$.
Below we argue that $\Gamma$ cannot vary much because the observed
peaks are often about the same width at the end of the burst as at the
beginning.  Thus, $\zeta$ is probably small.
For small $T_{0}$, this envelope is a pulse with a rise
that follows $\sim T^{2-4(\alpha+\zeta)}$ and
a fall that follows $T^{-\alpha-2}$.
Other functions for $P(T)$ can give sharper rises.  For example, if
the shell 
runs into some material such that $P(T)$ is a linearly increasing function
 from $T_{0}$ to $T_{\rm max}$, then the pulse can rise
as $\sim T^{3-4(\alpha+\zeta)}$.
If the falling density cause $P(T)$ to decrease with $T$ then
the pulse will rise slower than $T^{2-4(\alpha+\zeta)}$.

\centerline{\it 2.4 Infinitely Thin Shells Emitting at Multiple Radii}

Many bursts do not have a simple FRED-like shape but consist of many peaks.
One could have emission at multiple radii,
that is, have the shell emit at multiple times ($T_p$).
To investigate multiple radii, one can add functions like equation \VSINGLE~
with $T_0$ replaced with $T_p$.
A visual inspection of the BATSE catalog of multiple-peaked time
histories reveals
that usually peaks have about the same duration at the beginning
of the burst as near the end of the burst; that is, $\Delta T_p$ is
roughly constant for a burst
(see, e.g.,  Fig.~\FIGDIVERSITY c)
Equation (\VSINGLE) predicts that peaks should progressively become
wider
(i.e., the FWHM is $\propto$ $T_p$ and $T_p = T_0$ plus time from the
first gamma-ray peak).
  If peaks all originate with a
large $T_0$, the amount that the peaks widen will be smaller,
but then each peak must already be wide.  This is not observed (see
discussion of burst 219 in \S 4).
Note that $T/T_p$ = $t/t_p$ so the value of $\Gamma$ does not affect
how the peaks get progressively wider.

\centerline {\bf 3. RELATING PHYSICAL SIZE TO TEMPORAL STRUCTURE}

Except for FREDs, single infinitely thin shells do not predict
complex time histories (see \S2).
In this section, we
lift the assumption that the conversion of the bulk motion into
gamma-rays is locally spherically symmetric.
The thickness and the angular extent of the resulting emitting regions as well
as the duration of emission all contribute to form the time
histories.
  We will now look at five different scenarios
and how the observed time structure (e.g., pulse widths and
the burst duration) can be related to the physical size of the emitting region.

\centerline {\it 3.1 Simultaneous Conditions}

A shell expanding outwards could become photon-active,
turning on and off
because the appropriate
conditions for photon production occur at roughly the same time.
Such regions might occur in models of a shell sweeping up the ISM or
internal shock growth.
In this case,
  causality considerations do not limit the size of the
regions; a shell sweeping forward may become
photon-active at the same time in regions whose size is not limited by $ct'$.
  Let $\Delta r_{\perp}$ be the length of the
arc that becomes photon-active. Assuming for simplicity that the
active region is near the x-axis in Figure \FIGELLP, then the duration
of the pulse is reduced from $0.22 T_p$ to $\sim T_p
(\Delta r_{\perp}/2r_0)^2$.  The width is roughly the same for a patch
$\Gamma^{-1}$
away from the x-axis, as well.  
For example, assume that the arc responsible for
the peak in Figure \FIGVDELTA a is only active for a small range
of distances not near the x-axis.  The dotted lines in
Figure \FIGVDELTA a represent the possible peak shape if only a single patch
is active during the $\delta$-function emission.  The width of the
patch can be
adjusted to maintain a roughly constant peak width.
  If $\Delta T_p$ is the
width of the peak at time $T_p$ in the time history, then, for the
desired $\Delta T_p$ to be observed by the detector, $\Delta r_{\perp}$
must be limited to
\eqnam{\PATCHLIMIT}{PATCHLIMIT}
$$
\Delta r_{\perp} \lap 2 c \Gamma (T_p\Delta T_p)^{1/2}
\eqno(\neweq)
$$
This predicts that $\Delta T_p$ should scale as $\Delta r_{\perp}^2/T_p$
whereas often peaks have the same width throughout the burst.
Since there is no apparent reason why
 $\Delta r_{\perp}$ should scale as $T_p^{1/2}$,
it seems unlikely that the patch size, alone, 
gives the temporal width of the peaks.

\bigskip
3.1.1 Simultaneous Conditions and Burst Substructure

The angular size, $\Delta r_{\perp}$, and the width of each of the
emitting regions, $\Delta r_{\parallel}$, and the duration of emission
$\Delta t$ are all
constrained by the measured duration of the peaks, $\Delta T_p$.
The upper limits on each of these quantities, $\Delta r_{\perp},
\Delta r_{\parallel}, \Delta t$ are

$$
{}~~~~~~~~~~~\Delta r_{\perp, {\rm max}} = 2 \Gamma c (T_p\Delta T_p)^{1/2}
\eqno(\neweq a)
$$
$$
\Delta r_{\parallel, {\rm max}} = c \Delta T_p  
\eqno(\lasteq b)
$$
$$
\Delta t_{\rm max} = 2 \Gamma^2 \Delta T_p~.
\eqno(\lasteq c)
$$
We define three quantities, $\Delta T_{\Delta r_{\perp}}, \Delta
T_{\Delta r_{\parallel}}, \Delta T_{\Delta t}$, which are the
components of the peak time $\Delta T_p$ contributed by $\Delta
r_{\perp}$, $\Delta r_{\parallel}$, $\Delta t$, respectively.  Table 1
summarizes these quantities for each of the scenarios discussed.  For
a single shell that develops photon-emitting regions simultaneously,
these three quantities are
$$
{}~~~~~\Delta T_{\Delta r_{\perp}} = {\Delta r_{\perp}^2 \over 4 \Gamma^2
c^2 T_p}
\eqno(\neweq a)
$$
$$
\Delta T_{\Delta r_{\parallel}} = \Delta r_{\parallel}/c 
\eqno(\lasteq b)
$$
$$
\Delta T_{\Delta t} = \Delta t/ 2\Gamma^2~.
\eqno(\lasteq c)
$$

{}From equation (\lasteq) it is evident that $\Delta r_{\perp}$ must be
substantially larger than $\Delta r_{\parallel}$ for the curvature of
the expanding shell to have a significant effect on the time histories.
If the radius of emission were large (for example $T_0 \sim
100$ s) the patch spanned by $\Delta r_{\perp} = 2$ light seconds
would be
essentially a planar region causing a $\delta$-function emission in
the detector.  A $\Delta r_{\parallel}$ of 2 light seconds, however, will
cause a peak 2 s long in the detector.

\bigskip
3.1.2 Simultaneous Conditions and Burst Duration

The scaling of peaks within a burst does not generalize to the
scaling of the complete
\figname{\FIGEMITSHELL}{FIGEMITSHELL}
burst time histories.  Figure \FIGEMITSHELL~shows an example
of a mapping between the times of emission for various patches
in the detector rest frame and their arrival at the detector.  
The gamma-ray arrival at the detector is sensitive to
the angular offset from the line-of-sight at which the patch develops.
The dashed
lines represent the actual radii at which the patches reside.
The solid lines represent the locus of points from which emitted
photons arrive at the observer at the same time.  Any
difference in distance along the line-of-sight path, $\Delta d_{{\rm
los}}$, that the two
photons are emitted in the rest frame of the detector requires a time
$\Delta d_{los} / c$ to travel.  Thus even a small distance off axis
for a location of a patch can have a great effect on the time history.

Compare the photons emitted from site 2 to those from site 3.  Site 2
photons were emitted at a smaller radius and thus earlier than site 3,
but will arrive later because of their angular offset.  Likewise,
those emitted from
site 4 and site 5 will arrive at the observer at the same time, even
though site 4 photons were emitted significantly earlier.  The degree
by which the peaks may be scrambled can be seen in Figure \FIGELLP.  The
region of the ellipsoid which is far
from the line-of-sight corresponds to photons which 
were emitted at $T/2$ earlier than
photons emitted along the line-of-sight.  Thus the order of the peaks
can be
scrambled by $\sim T/2$.

The order of arrival of the peaks is a function of both the
time of emission and angular offset and it
is not possible to solve this inverse problem of peak order in the
CMF (or the DRF) with certainty.

One conclusion which may be drawn from this analysis is that a shell
cannot decelerate significantly during photon emission, otherwise a
noticeable evolution in the time history would occur.
 In Figure \FIGDIVERSITY, the
$\Delta T_{p}$'s in BATSE burst 2856 remain relatively constant
throughout the burst.  Since $\Delta T_p \propto \Delta t/\Gamma^2$,
either there is a negative correlation between $\Delta t$ and
$\Gamma^2$ or $\Delta t$ and $\Gamma^2$
do not change significantly throughout the burst.
If the widths of peaks in any given GRB do
not change by more than a factor of 3 or 4, $\Gamma$ must not change
by more than a factor of 2 within a given burst.  Norris et al. (1996) reports
that peaks vary only from 10 ms to 2 s in  bright bursts,
implying that $\Gamma$ cannot change by more than $10^{5/2}$ from burst to
burst in all scenarios where $\Delta t \propto \Gamma^{-2}$.

\centerline {\it 3.2 Perturbation Growth}

In this second scenario, a gamma-emitting region grows from a seed in the
CMF at
 speed $c_s'$.
In contrast to \S3.2, the source size is limited by causality because the
perturbation grows at a finite signal speed.
  Woods \& Loeb (1995) discuss such growth, although only on
the surface of the shell rather than in three
dimensions within the shell.
We first examine seed growth in the context of forming the
substructure of a burst and then consider how multiple seeds can form
the complete time history of a GRB.

\bigskip
3.2.1 Perturbation Growth and Burst Substructure

Suppose that a perturbation grows  at speed
$c_s'$ for a time $\Delta t'$ in the comoving frame.  After a time
$\Delta t'$, the perturbation has grown in three dimensions to the
size
\eqnam{\SGRPERP}{SGRPERP}
$$
\Delta r_{\perp}' \sim \Delta r_{\parallel}' \sim c_s' \Delta t'
\eqno(\neweq)
$$
The gamma-rays from this region will be emitted over both a range of
times, spanning $\Delta t'$, and a range of radii, $\Delta
r'_{\parallel}$, along
the line-of-sight toward the detector.  We look at how these effects
couple to determine the mapping from the DRF to the detector's arrival time.

The duration of the seed growth will map in the same manner as
duration has mapped in all situations (cf. Fig.~\FIGTHICKDUR~and \S 2.1):
\eqnam{\SGTD}{SGTD}
$$
\Delta T_{\Delta t} = {1 \over 2 \Gamma^2} \Delta t = {1 \over 2 \Gamma}
\Delta t'
\eqno(\neweq)
$$

The thickness of the
perturbation in the DRF will
determine the time difference between two photons emitted at the same
time in the DRF at the front and back of the shell
(cf. Fig.~\FIGTHICKDUR).  The
maximum thickness of a patch which grows for a comoving time $\Delta t'$ at
speed $c_s'$ is
\eqnam{\SGRPAR}{SGRPAR}
$$
\Delta r_{\parallel} = \Delta r_{\parallel}' \Gamma^{-1} = c_s' \Delta
t' \Gamma^{-1} = c_s' \Delta t \Gamma^{-2}~.
\eqno(\neweq)
$$
Thus, the maximum time spread due to thickness is
\eqnam{\SGTTH}{SGTTH}
$$
\Delta T_{\Delta r_{\parallel}} = 
{\Delta r_{\parallel} \over c} = {c_s' \over c}
\Delta t \Gamma^{-2}~.
\eqno(\neweq)
$$
In this scenario, $\Delta r_{\perp}' \sim \Delta r_{\parallel}'$, which
leads to
\eqnam{\SGRPERP}{SGRPERP}
$$
\Delta T_{\Delta r_{\perp}} = {c_s'^2 \Delta t^2 \over 4 \Gamma^6 c^2
T_p}
\eqno(\neweq)
$$
which we can neglect.

We assume that the timescales from $\Delta T_{\Delta r_{\parallel}}$
and $\Delta T_{\Delta t}$ add in
quadrature to produce the observed peak:
\eqnam{\SGTOB}{SGTOB}
$$
\Delta T_p = {\Delta t \over \Gamma^2}((1/2)^2 + (c_s'/c)^2)^{1/2}
\eqno(\neweq)
$$

\bigskip

3.2.2 Perturbation Growth and Burst Duration

The perturbations occurring in different locations on the expanding
shell emit peaks whose order of arrival at the detector is affected by
the angular offset from the line-of-sight to the detector in much the
same way as those from simultaneous patches 
(cf. \S 3.1.2).  Again, this offset can
scramble the arrival of the peaks by as much as $T_p/2$ in which case
any time evolution of the morphology of the peaks of photons emitted
would still be apparent, although more difficult to recognize.

\centerline {\it 3.3 Ambient Objects}

\def\DRamb{\Delta r_{\rm amb}}
\def\DRpar{\Delta r_{\parallel}}
\def\DRperp{\Delta r_{\perp}}
In the previous sections, the time structure arose from variations
within the expanding shell.  In this section, it is assumed that the
time variation occurs because the shell interacts with inhomogeneities
within the ambient medium.   The shells might
encounter clouds with variations in density (Rees and \Mesz 1994).
Or, the shell might sweep over a source of photons, up scattering those ambient
photons to gamma-rays.
For example, Epstein et al. (1993) pointed out that
a typical GRB spectrum looks like an AGN spectrum that has
been boosted.  Furthermore, an AGN is one of the few places that has enough
photons to flood the rest of the universe.  Time variations might arise
because of variations in the brightness of blobs ejected from the AGN
(Epstein et al.~1993).
A key problem with this suggestion  is that well
localized GRBs do not seem to be correlated with
the luminous mass (the ``no host'' problem,
Fenimore et al.~1993b).
Another suggestion is that the bursts occur in the collapsed cores of
globular clusters at
cosmological distances and each peak is due to
the shell running over an individual star (Shemi 1994; Shaviv \& Dar 1995).
This scenario does not work because it assumes that most of the peaks
are due to stars at angles relative to the line-of-sight that are
much larger than $\Gamma^{-1}$.  We will ignore the issue of whether
there would be sufficient photons at such large beaming angles.  The
central problem is that Shaviv \& Dar (1995) assumed that the rest frame time
and the arrival time always scales as $(2\Gamma^2)^{-1}$ at all angles.
In fact, that is only valid for small angles; the actual relationship
is $1-\beta \cos \theta$.
At larger angles which include enough stars to account for the peaks,
the peaks would spread out over $\sim 10^6$ s in arrival time.

3.3.1 Ambient Objects and Burst Substructure

Let us call the thickness of the expanding shell $\Delta r_{\parallel}$
and the radius of the ambient object in the rest frame $\Delta r_{\rm amb}$.
Whether or not the ambient object collapses onto the shell
is a key distinction that must
be made in order to understand how the ambient object determines the
time structure.
For example, a shell interacting with a cloud might be expected to sweep
up the material, effectively collapsing the cloud onto the surface of the
shell.
In this case, the photon production is similar to that of photon ``A'' and
``a'' in Figure \FIGTHICKDUR.  
The shell keeps up with the photons it produces and
the duration of the collapse appears in arrival time as:
\eqnam{\PFTT}{PFTT}
$$
\Delta T_{\Delta t} = {\Delta r_{\rm amb} \over 2 \Gamma^2 c}~~.
\eqno(\neweq)
$$
The ambient object is most likely symmetric so the size in the
perpendicular direction is the same as in the parallel direction.
Thus, there will also be a contribution
\eqnam{\PFTRPERP}{PFTRPERP}
$$
\Delta T_{\Delta r_{\perp}} =
{\DRamb^2 \over 4\Gamma^2 c^2 T_p}~.
\eqno(\neweq)
$$
Since $\Delta r_{\parallel}$ is zero for collapsible objects (by definition),
the observed peak width will be some combination of $\Delta T_{\Delta T}$
and $\Delta T_{\Delta r_{\parallel}}$.
Unless $T_p \gap \Gamma^2 \Delta T_p$ (unlikely), the observed width
will be dominated by $\Delta T_{\Delta r_{\parallel}}$ and
$\DRamb$ $\lap$ $2\Gamma c(T_p\Delta T_p)^{1/2}$.
The observed widths of peaks often do not scale 
as $T_p^{-1}$, thus, it seems that
collapsible objects would have a difficult time explaining the time history.

The alternative is that the ambient source does not collapse, but produces
gamma-rays on a scale of $\Delta r_{\rm amb}$.
For example, stars would not collapse.
In this case, the effect of $\DRamb$ scales as a thickness
much like photons ``A'' and ``B'' in Figure \FIGTHICKDUR~
and the width is determined from the
 light travel time across the overlap of the shell thickness
and the ambient source thickness.
Both $\Delta T_{\Delta T}$ and $\Delta T_{\Delta r_{\perp}}$ are small, so
\eqnam{\PFTOB}{PFTOB}
$$
\Delta T_p \sim \Delta T_{\Delta r_{\parallel}} =
{{\rm min}(\Delta r_{\parallel}, \Delta r_{\rm amb}) \over c}~.
\eqno(\neweq)
$$

  We note that usually all of the peaks appear to
be about the same size (see Fig.~\FIGDIVERSITY c and 7c). 
 Stars might have similar sizes
 but there is an insufficient density of stars to account for the time
history (see below).  If the width of the peaks were due to the
distribution of cloud sizes, one might expect a power law distribution of
peak widths, which is not observed.
Thus, it appears that $\Delta r_{\parallel}$ is smaller than
$\Delta r_{\rm amb}$ and determines $\Delta T_p$.
An upper limit on $\DRamb$ is set by the
condition that $\Delta T_{\Delta r_\perp}$
does not dominate.
Therefore,
\eqnam{\RAMBLIMIT}{RAMBLIMIT}
$$
3 \times 10^{10} {\rm cm} = c\Delta T_p < \Delta r_{\rm amb} < 2\Gamma c
(\Delta T_p~T_p)^{1/2} \sim 3 \times 10^{13}\GSCALE~~ {\rm cm}~
\eqno(\neweq)
$$
where $\GSCALE$ is $10^{-2}\Gamma$.

\bigskip
3.3.2 Ambient Objects and Burst Duration

The duration of events defined by ambient objects 
depends on their distribution.
The ambient objects responsible for the peaks are contained in the
volume swept out by the shell within $\theta \sim \Gamma^{-1}$.
This volume is a cone with a height of $2\Gamma^2cT_p$ and a base radius
of $\Gamma cT_p$.  The volume grows as $T_p^3$ so one might expect the number
of peaks per unit time to grow as $T^2$ (which, of course, is not observed).
The shell apparently cannot utilize the ambient objects until after
some time $T_0$ perhaps because some ISM material must first be swept up
before the shell can interact with the ambient objects  to generate gamma-rays.
If the burst duration is $T_b$ and $N_b$ peaks are observed, the
required density is
\eqnam{\RDENSITY}{RDENSITY}
$$
\rho_{\rm amb} \sim 
{N_b \over {2 \over 3} \pi \Gamma^4 c^3 \big((T_0+T_b)^3-T_0^3\big)}~.
\eqno(\neweq)
$$
We have clear limits on $T_0$: $T_0$ cannot be much smaller than $T_b$
because we usually do not see the number of peaks increase as $T^2$.
Furthermore, $T_0$ cannot be much larger than $T_b/0.22$ since otherwise
by local spherical symmetry, the
objects at angles $\sim \Gamma^{-1}$ would produce peaks later than observed.
Using $T_0 \sim (5/2)T_b$,
\eqnam{\RDEN}{RDEN}
$$
\rho \sim 1.4 \times 10^8 N_b \GSCALE^{-4}(100/T_b)^3~~{\rm pc}^{-3}~.
\eqno(\neweq)
$$
We have treated $\Gamma$ as a free parameter.  Often, one estimates that
the photon-quiet phase is roughly the time it takes to energized the swept-up
ISM, that is (\Mesz \& Rees 1993),
\eqnam{\SWEPTUP}{SWEPTUP}
$$
E_0 \sim {4\pi \over 3} r_0^3 \rho_{\rm ISM} m_pc^2\Gamma^2
\eqno(\neweq)
$$
where $E_0$ is the total radiated energy, $\rho_{\rm ISM}$ is the density of
the ISM (typically, 1 cm$^{-3}$) and $m_p$ is the mass of a proton.
We have shown that, due to shell symmetry, the radius of the photon-quiet
phase is roughly $r_0 \sim 2\Gamma^2 cT_{1/2}/0.22$ where $T_{1/2}$ is the
FWHM. Therefore,
\eqnam{\E0LIMIT}{E0LIMIT}
$$
E_0 \sim 1.3 \times 10^{32} T_{1/2}^3 \Gamma^8
\eqno(\neweq)
$$
A reasonable value of $E_0$ is $10^{51}$ erg.
Certainly, $\Gamma \sim 10^3$ is not allowed since it would require
$E_0 \sim 10^{58}$ erg.  Thus, $\Gamma$ must be the order of 50 to 100.
This justifies the scaling of $\Gamma$ in equation (\RDEN).

Since $T_p \sim T_0 + T_b$,
the size of the region is $\sim
 2\Gamma^2(T_0+T_b) \sim
 0.07 \GSCALE^{2}(T_b/100)$ pc.
For burst 2856, $N_b \gap 75$ and $T_b \sim 150$ s, so $\rho$ must
be the order of $3 \times 10^{9}$
 pc$^{-3}$
within a radius of 0.1 pc
 if $\Gamma = 100$.
Certainly, this density is too high to be achieved by stars.

\centerline{\it 3.4 Thick Shell with Substructure}

Consider the situation where material is ejected from the central
site with a range of Lorentz factors.
If the velocities of the emitted particles vary during the release time, a
rough differentiation of velocities will occur and the expanding shell
will spread to a finite thickness (Rees \& \Mesz 1994, Piran 1994).
Suppose that the front of the shell moves at $\Gamma_{\rm max}$ and the
back at $\Gamma_{\rm min}$.
Let $T_b$ be the duration of the event as measured in the arrival time.
  Once the shell reaches a thickness
of $\sim cT_b$, the time history
can be formed simply by varying the emission of photons in spatial
coordinates in the DRF; emission from a region with $\Delta
r_{\parallel} = cT_b$ will map to a time history of duration $T_b$ in
\figname{\FIGTHICKSUB}{FIGTHICKSUB}
the detector (see Fig.~\FIGTHICKSUB).
Each subpeak arises from a region the order of $c\Delta T_p$.
Alternatively, a central site emitting for a duration $T_b$ 
could produce a thick shell
with substructure even without a spread in $\Gamma$ (Piran 1994).

To be relevant, the thick shell must
 grow to a $\Delta r_{\parallel} = cT_b$ within a radius
$r_{\rm max} = 2\Gamma_{\rm max}^2c(T_0+T_b)$.
Otherwise, the duration is dominated by the shell curvature 
(i.e., $T_b \sim 0.44T_0$).
In Figure \FIGTHICKSUB, the vertical dashed line subtends $\sim 2\Gamma^{-1}$.
Let $d_c$ be the curvature within $\sim 2\Gamma^{-1}$.
If the shell is much thicker than $d_c$, the observed time structure can
arise from spatial substructure within the thick shell.
To be thick enough,
$\Gamma_{\rm max}$ and
$\Gamma_{\rm min}$ are related by
\eqnam{\TSGAMMA}{TSGAMMA}
$$
(v_{\rm max} - v_{\rm min})2\Gamma^2_{\rm max}(T_0+T_b) = 
\bigl({\Gamma_{\rm max}^2 \over
\Gamma_{\rm min}^2} - 1 \bigr)c(T_0+T_b)
>0.44(T_0+T_b)
\eqno(\neweq)
$$
which leads to 
$$
\Gamma_{\rm max} > \sqrt{1.4}\Gamma_{\rm min}~.
\eqno(\neweq)
$$
 Rees \& \Mesz (1994) discuss shocks with a
$\Gamma_{\rm max} / \Gamma_{\rm min} = 2$ in which shells of varying
$\Gamma$'s collide and form shocks which emit gamma-rays.
In that model, some slow speed material leaves the central site first such that
some high speed material catches up and causes shocks.  In our scenario,
we emphasize that
 the high
speed material leaves first and stretches out the thickness of the shell.
This effectively converts the shell into a parallel slab and most of the
problems associated with local spherical symmetry are eliminated.

\centerline{\it 3.5 Central Engine}

Unlike the previous scenarios, a pure central engine does not involve a single,
expanding shell which moves away from the central site.  Rather,
the central engine emits multiple shells for a period 
roughly equal to the duration of
the observed GRBs (up to $\sim 10^3$ s).
The shells expand to a radius of $\sim 2\Gamma^2c\Delta T_p$.
  A
release of these shells over a duration $T_b$ in the DRF (which
is also the rest frame of the central site), results in the
arrival of these shells in the detector over the same duration $T_b$.
In this scenario, the radius of the shell does not exceed that
given by the duration of the peaks in the time history.

\centerline{\it 3.6 Summary of Relationships Between Temporal Variations}
\centerline{\it and the Physical Size}

As the previous sections demonstrate, there are many ways
to interpret the observed time histories. Table 2 summaries these.
Here, $\Delta T_p$ is a typical pulse width, $T_b$ is the observed duration
of the event, $T_0$ is the duration of the photon quiet phase, and
$T_D$ is the total duration of the event ($=T_0+T_b$).
In Table 2, $R$ is the scale of the entire event and $\Delta R_{\perp}$ is the
perpendicular scale of the emitting region.
The $\Delta R_{\parallel}$ is always less than 
$c\Delta T_p$ since thickness effects
are not affected by the expansion (see \S2.1 and Fig.~\FIGTHICKDUR).
A typical use of $R$ would be to estimate the amount of ISM material that the
event sweeps up.  A typical use of $\Delta R_{\perp}$ and
$\Delta R_{\parallel}$ is to estimate the
photon density in calculations of photon-photon attenuation.
For example, Fenimore, Epstein \& Ho (1993a) explicitly assumed
either a stationary photosphere or
a central engine
in estimating  the size of the region relevant
for calculating the photon-photon attenuation.
Nayakshin \& Fenimore (1996) use the results of this paper to obtain
a more realistic estimate.

In the first three cases in Table 2, $R$ arises from the size of the
single shell (cf.~eq.~[\TTREL]).
Patches from simultaneous conditions are limited by equation
\PATCHLIMIT~whereas patches that are limited by causality (i.e., growth
from a seed) are limited by equation
\SGTOB.
Thick shells with substructure are dominated by thickness effects so the
duration arises from the overall thickness of the shell grows and
the individual emitting regions are set by $\Delta T_p$.
The thick shell with substructure could arise from either a single
shell that spreads out due to a spread in $\Gamma$ (cf.~\S3.4) or by
a central engine that continuously feeds a shell.
 In the latter case, $R$ is limited to
to $cT_b$.
Finally, for completeness, we include a stationary photosphere with a
relativistic wind (cf.~\Pacz 1986, 1990).  Since there is no motion of the
emitting surface, the size is limited by the light travel across the observable
surface (i.e., the surface within $\Gamma^{-1}$).

Although at first glance, many of the scenarios 
that we have discussed seem to be very
similar, Table 2 demonstrates that one obtains different estimates of the
overall size or the size of the emitting region in each case.
 The overall size of the region can
vary from $cT_D$ to $2\Gamma^2T_D$ and the perpendicular size of the emitting
region can vary from $c\Delta T_p$ to $2c\Gamma T_p$.
We have explicitly only considered kinematic limits and have not considered
the fireball physics that might cause these changes.  Several types of
fireball models have been considered in the literature.  Some
involve ``impulsive'' fireballs where the time structure arises in the
expanding shell.  In Table 2, the impulsive models include the thin shell,
patches from simultaneous conditions (probably caused by ISM variations, see
\Mesz \& Rees, 1993), patches from seeds (probably caused by internal
shocks, see Rees \& \Mesz 1994, Piran 1994, Woods \& Loeb 1995), and the
multiple shell scenario.  Other models involve ``continuous energy
input fireballs''.  In Table 2, these models include the thick shell
scenarios (see Rees \&
\Mesz 1994, Waxman \& Piran 1994) and the stationary photosphere (\Pacz 1990).
The thick shell from a spread in $\Gamma$ is a hybrid where the time structure
might arise from the duration of the energy input or
the spread in size after the impulsive release.

\centerline{\bf 4. DISCUSSION}

We have outlined five general scenarios in which most
existing models of GRB emission can fall.  Each of these
scenarios has constraints outlined in Table 1 on photon emission based
on the observations.  We now turn to three general classifications
covering the majority of burst types and discuss how the burst scenarios
can be used to account for GRB time structure.  The three
classifications are bursts of
single or a few isolated sharp spikes, FREDs, and bursts with complex
time structure.

\centerline{\it 4.1 Sharp, Isolated Spikes}
  The time histories of bursts which fall
into this category consist of one or, at most, a few sharp spikes
occurring within a short duration.
An example of such a burst is shown 
in Figure \FIGDIVERSITY a.  Note that the fall time
is roughly the  same as the rise time so this is not just a very
narrow FRED.
  Such spikes can be explained by
one of three scenarios.
First,
  the emission may arise from a full shell
with local spherical symmetry
emitting photons at a low $t_0$ so that the FWHM due to the shell ($\sim
0.22  t_0/(2\Gamma^2)$) is smaller than the other contributors
to the width.
For example, consider the contribution arising from how the gamma-emitting
region could grow from a seed over time $\Delta t$.  The lack of a
FRED-like shape in Figure \FIGDIVERSITY a implies that $t_0 \lap \Delta t/2$.
Since $\Delta T_p$ of Figure \FIGDIVERSITY a is only $\sim 0.5$ s, $t_0$ must
be $\lap 5\Gamma^2$, or $5 \times 10^4$ s for $\Gamma = 100$.

Second,
the photons which cause the single spike 
could arise from a single localized patch or
seed at a larger $t_0$.
Although the bulk material might expand under local spherical symmetry,
the conversion to gamma-rays could be very asymmetric.
  In this case $r_0$ is not limited to being
small, because $\Delta r_{\perp}$ is assumed to be small to produce the
sharp time profiles observed.  The region must, however, be
significantly smaller than the total region that lies within the beam
toward the detector.  If the photons we observe were emitted from
only a small portion of the shell, then the total energy requirements
of the shell becomes greater by the ratio of the total surface to
that of the emitting surface.
We define $f$ to be the
filling factor, that is the ratio of the observed emission
to that which would be expected under local spherical symmetry of the
gamma-ray production:
\eqnam{\FILLFACT}{FILLFACT}
$$
f \propto  {\int P(\theta,\phi,t)\Lambda^{-3}
                  \,dA  \over
           \int\Lambda^{-3}
                  \,dA   }~~.
\eqno(\neweq)
$$
For a patch to give a time scale of only $\Delta T_p$, the filling factor
needs to be:
\eqnam{\ENERGY}{ENERGY}
$$
f  \propto
{ r_{\perp}^2 \over r^2 \Gamma^2} =
 { \Delta T_p \over
 T_p}
= {\Delta t_p \over t_p}~.
\eqno(\neweq)
$$
The shell would have produced a FRED if all of it emitted but only a small
region (e.g., patch 5 in Fig.~\FIGEMITSHELL) emitted resulting in the
dotted line in Figure \FIGVDELTA a.
It is normally assumed that the fireball energy 
is first converted to kinetic energy of the shell
and then converted to gamma-rays.
To explain narrow spikes if $t_0$ is larger than $\sim 10^4$, one must
assume either that the  material is confined to a pencil beam with an opening
angle of only $\sim \Delta t_p/t_0$ or that only a region with a size
$\sim\Delta t_p/t_0$ converted its energy into gamma-rays.
In the latter case, $f$ is very small and most of the energy of the
fireball is never converted into gamma-rays.
This could raise the required energy to larger than that available by
merging neutron stars.

Third,
a shell sweeping past a star (\S 3.3) can account for single sharp
spikes.
However, since the star only interacts with a small fraction of the shell's
surface ($\sim \Delta r_{\rm amb}/\Gamma c t_0$), the filling factor ($f$)
is, again, small.

\centerline{\it 4.2 FREDs}

FREDs account for a significant proportion of all GRB time
histories.  They can be explained easily by a single, full shell
emitting for a short $\,dt$ at a radius $r_0 = 2 \Gamma^2 c T_0$, where
$0.22 T_0 = \Delta T_p$.  This $\delta$-function emission discussed in \S
2 produces an envelope very similar to many observed FREDs.
  It has a sharp rise ending in a cusp and falls off slowly
to background. No other scenario can explain so simply the
morphology of FRED GRBs.
The FRED shape in arrival time is determined solely from $T_p$.
For example, we have fit equation (\VSINGLE) to the time history of burst
1885 (see Fig.~\FIGDIVERSITY b).  
The FWHM of the FRED is about 35 s so $T_0$ is 157 s.
We have placed $T = 0$ in Figure \FIGDIVERSITY b about 157 s before the FRED.
{}From equation (\E0LIMIT), one can estimate $\Gamma$.  Even
using a large value
of $E_0$ ($10^{52}$ erg), $T_{1/2}$ equal to 35 s implies that $\Gamma < 80$.

\figname{\FIGFITEX}{FIGFITEX}
Figure \FIGFITEX a shows burst 678 which has statistically significant subpeaks
within an envelope that is roughly FRED-like.
The $T_{1/2}$ is 7.7 s so equation (\E0LIMIT) implies $\Gamma \sim 140$.
The presence of the subpeaks
is easily explained as separate regions that became active roughly
simultaneously in the DRF.
If the conversion of bulk motion to gamma-rays has local spherical symmetry
on average over the shell,
some regions could be more efficient at producing gamma-rays than others.
The peaks in the tail of the FRED arrive later 
at the detector because they are due
to regions offset from the line connecting the central explosion and the
observer.
Since the late arriving peaks are due to regions offset by angle $\theta$
from the tip of the ellipsoid, those regions should have
a different Lorentz boost factor.
The region near the tip of the ellipsoid produces the signal near the
peak of the FRED and
its spectrum is the comoving spectrum boosted by $\Lambda_p^{-1} = 2\Gamma$
(see eq.~[\SCV]).
For a peak arriving at time $T$, the boost factor is $2\Gamma T_p/T$.
Thus, we predict that the relative boost factor in a FRED is simply
$T_p/T$.
Note that one needs to know where $T = 0$ is but that can be found from the
shape of the FRED.
  There is a one-to-one mapping of the relative boost
factor to the FRED shape; this is shown is Figure \FIGVDELTA b.
In Figure \FIGFITEX a, the arrow indicates 
where the spectrum should be boosted by
a factor of 2 less than at the peak of the FRED
(assuming $\alpha = 1.5$).
Indeed, burst 678 and other FREDs show substantial softening 
through the tail.  A detailed
analysis to see if the softening is consistent 
with Figure \FIGVDELTA b is in preparation.
Since time is also affected by the Lorentz transformation, the average
temporal structure at the arrow should be twice as long as the temporal
structure near the peak.

In contrast to burst 678, some GRBs (e.g., burst 451) have extremely
smooth profiles where the intensity follows a clear pattern.
Smooth-profile GRBs could present special problems for relativistic shells.
In the context of relativistic shells, the smoothness arises because
of nearly identical conditions on a shell over distances the order of
$\sim \Gamma cT$.
Causality is not violated because the coordinated brightness comes about from
local spherical symmetry, not by the propagation of a faster-than-light
signal.
However, if the smoothness of the intensity is from nearly identical
conditions, than the spectral softening should be expected to closely
follow $\Lambda/\Lambda_p = T_p/T$.
If it does not follow $\Lambda/\Lambda_p$, then the smoothness probably arise
from the propagation of a signal and the perpendicular  size
is limited by causality
($c\Gamma \Delta T_p$) rather
than by the shell size
($2c\Gamma T_D$, see Table 2).
In such a case, the filling factor is very small ($\sim[\Delta T_p/T_D]^{2}$)
and the gamma-ray producing regions must
subtend only $\sim(\Delta T_p/T_D/\Gamma)^{2}$ sterradians.
Since the region is small, the required $\Gamma$ to avoid strong
photon-photon attenuation is larger (Nayakshin \& Fenimore 1996).
The net result is that the emission would be confined to angular scales of
$<10^{-8}$ sterradians, a serious problem for any cosmological model.
Thus, if the smooth-profile GRBs do not have the predicted softening pattern,
the emitting region is probably unacceptably small.

Figure \FIGFITEX b shows
burst 219 which has several large peaks on a FRED-like envelope.
Thus, it is similar to burst 678 in Figure \FIGFITEX a except that
some patches appear to have a wider range of 
 gamma-ray efficiency producing a wider
variation in peak height.
The dashed curve is the widest FRED-like shape that we could fit to
the envelope of emission.
Including more of the large peaks in the fit would make the FWHM of the
FRED-like shape smaller.  Based on the FWHM, the time equal to zero in Figure
\FIGFITEX b is when the shell responsible for the emission started.
We believe that it is reasonable to attribute the peak shape to an
expanding relativistic shell that became gamma-active after
$T_0 \sim 30$ s.
However, the peak shown in Figure \FIGFITEX b is actually the second peak from
this event.  The first peak occurred 100 s earlier.
If one assumes a single explosion (e.g., two merging neutron stars),
$T_p$ must be at least 100 s for the second peak.
The solid curve in Figure \FIGFITEX b shows the expected FRED-like shape
based on $T_p$ being at least 100 s.
Either the two peaks in burst 219 have a temporal separation because they were
caused by two separate explosions (i.e., a central engine) or the second
peak must have a  low $f$, and only a small portion
of the shell within $\Gamma^{-1}$ became gamma-active.
We have used $\zeta$ = 0 (cf.~eq.~[\POWERINDEX]).
  If $\zeta > 0$, the discrepancy between the
observed envelope and the predicted from the precursor would even be
larger.

We consider burst 219 to be an excellent example 
which points to central engines
as the origin of the overall duration of GRBs.
Koshut et al.~(1995) define precursor activity as weak emission that is
separated from the remaining emission by an interval that is at least as long
as the remaining emission.
Using the notation of Koshut et al.~(1995), any burst that has $\tau_{\rm main}
{}~\gap~ 2.5 \Delta t_{\rm det}$ is a strong contender for being a central
engine.
(Here, one should use an estimate of the duration of the FRED-like
component for $\tau_{\rm main}$.)~~
A few percent of the BATSE bursts probably satisfy that condition.
A significant difficulty which lies in a central engine explanation
is that cosmological objects must emit $\sim
10^{50}$ erg on a time scale that compares to the observed durations.

\centerline{\it 4.3 Complex Bursts}

  This category includes all bursts whose time
histories contain more substructure than those discussed above.
  The bursts in Figures \FIGDIVERSITY c and \FIGFITEX c
fall into this category.
We will discuss three scenarios that could produce long complex bursts.
The first is trivial: a central engine can explain any time history.
The second involves expanding shells which convert their energy into gamma-rays
because of some process internal to the shell (such as shocks).
The third scenario involves interactions with ambient objects external
to the shell such as clouds or stars.

The basic problem with attempting to explain 
large complex bursts with shells that
expand and then convert their kinetic energy into gamma-rays is that it
should be possible to discern a FRED-like shape or a sum of FRED-like
shapes.
In Figure \FIGFITEX c we have attempted to identify ways that $V(T)$ could
account for the time history.  A single shell (e.g., the dashed curve
in Fig.~\FIGFITEX c) implies low $f$: many portions of the shell did
not convert their energy into gamma-rays.
Multiple shells could help increase $f$.  The five solid lines
represent shells at five different $T_p$'s.
Note that each successive shell produces a FRED-like structure
that is progressively wider.
Again, $f$ would be small.
Another alternative is a shell that emits for a range of times.
The dotted line in Figure \FIGFITEX c is a fit using equation (\VTHICK).
Again, the quiet times have to be attributed to regions that do not
turn on as gamma-ray sources implying a low $f$.

The third scenario that might be applicable to long complex bursts is
one in which
the time structure comes from the shell interacting with ambient
inhomogeneities such as clouds in the ISM or stars.
Nowhere is the density of stars high enough to provide the inhomogeneities.
Collapsible objects (see \S 3.3) should produce peaks that have variations
in their widths (see $\Delta T_{\Delta r_{\perp}}$ in Table 1).
Long complex bursts often have peaks which are remarkably similar throughout
the burst (see Fig.~\FIGDIVERSITY c).
One expects $\Gamma$ to vary as the shell loses energy and there is no
reason why all the ambient objects should be the same size.
  Only peaks from non-collapsible objects are independent
of $\Gamma$ and the size distribution of the objects.
Ambient objects are acceptable only if the
 objects are non-collapsible, 
are between $3 \times 10^{10}$ and $3 \times 10^{13}$ cm
in size (eq.~[\RAMBLIMIT]),
and the density
is the order of $3 \times 10^{9} (\Gamma/100)^{-4}$
pc$^{-3}$
over a region of $\sim 0.1\GSCALE$ pc (eq.~[\RDEN]).
It is not clear what ambient objects can fulfill these characteristics.
Stars have the right size and are non-collapsible but do not have
the requisite density.
ISM variations are probably collapsible and they probably usually have
larger sizes.
Previously, it was thought that $T_0$ could be very large reducing
the required density.  However, large $T_0$ cannot be much bigger than
the observed duration since otherwise additional peaks would be seen at
$\sim 0.22T_0$.  Perhaps $\Gamma$ is as large as $10^3$.  Then
the required objects can be as large as $10^{14}$ cm, the density as
small as $3 \times 10^{6}$ pc$^{-3}$ within a radius of 10 pc.
However, then equation (\E0LIMIT) requires $E_0$ to be $10^{62}$ erg.
  In any case, these ambient objects cannot be arranged
randomly since the separation in the peaks occurs with a
log-normal distribution rather than with a random
distribution (Li \& Fenimore~1996).

Thus, from the kinematics and the observed time structure, we conclude that
it is very difficult to explain long complex bursts 
with a single shell (except for FREDs).
It seems that such structure can only arise from either
a central engine or
a thick shell with substructure (e.g., Rees \& \Mesz 1994, Piran 1994,
Waxman \& Piran 1994).

\centerline{\bf 5.~CONCLUSIONS}

The large Lorentz factor required to allow the high-energy photons escape
from GRBs results in a shell that is only visible when seen head on, GRBs
rarely appear as the sides of a jet.  When viewed head on, the curvature
of the shell is just as important as the expansion in determining the
temporal structure.
Thus, local spherical symmetry needs to be assumed.
The principle of local spherical symmetry can easily explain the
FRED-like bursts.
We predict that, if FREDs are the result of a shell, they can
be fit by $T_0(T/T_0)^{-\alpha-2}$ where the only true free parameter
is the start time.
The spectra should soften as $T/T_0$. FRED-like bursts with
precursors provide a strong argument that the time histories are
due to a central engine rather than a single release of energy 
(see Fig.~\FIGFITEX b).

If smooth-profile bursts do not soften as $T/T_0$, then the size of the
emitting region is probably determined by light-travel time and, therefore,
extremely small: $<(\Delta T_p/T_D/\Gamma)^{2}~\sim 10^{-8}$ sterradians.

Short spikes must have very small photon-quiet phases ($10^{4}$ s), or the
material is beamed on a scale much smaller than $\Gamma^{-1}$, or the
conversion of bulk motion to gamma-rays must have a small filling
factor, that is, only a small portion of the shell is ever induced into
producing gamma-rays.  This would raise the overall energy requirements.

 Long complex bursts present a myriad of problems for the models.
The duration of the event is $\sim t_0/(2\Gamma^2)$.
The long duration cannot be due
to large $t_0$ since it requires too much energy to sweep up the
ISM (eq.~[\E0LIMIT]).
Nor can it be due to
small $\Gamma$ if the time variation is due to
ambient objects since the density of such objects 
is unreasonable (Eq.~[\RDEN]).
Long events with small $\Gamma$ and time variations due to shocks must
explain why they almost always violate local spherical symmetry.
  We introduce the ``Shell
Symmetry'' problem for cosmological GRBs: models that arise from a
single, central release of energy that forms a relativistic shell must somehow
explain how either the material is confined to pencil beams narrower
than $\Gamma^{-1}$ or how a shell can have a low filling factor
with the resulting higher energy requirements.
Without such explanations, local spherical symmetry requires a FRED-like
shape.
Ambient objects can be inhomogeneous and there could be situations where they
would produce a low filling factor.  However, this might be able to explain
a few non-FRED time histories whereas most GRB time histories are non-FRED.
  We show that the thickness (in contrast to the duration)
of the gamma-ray emission can easily dominate because the contribution
from the thickness is not affected by $\Gamma$ (see Table 1).
Many bursts have 1 s time scales so the emitting region cannot be thicker
than 1 s whereas the radius of the shell is $2\Gamma^2T$ where $T$
can be hundreds of seconds.
The number of peaks per unit time does not usually increase as $T^2$,
thus $T_0$ must be large.  However, a large $T_0$ should produce a FRED-like
envelope with a width of $\sim 0.22T_0$.
The width of the peaks can remain remarkably constant throughout a long burst.
Only ``non-collapsible'' ambient objects produces peak widths independent
of $\Gamma$ and the size distribution of the ambient objects because
non-collapsible objects produce peaks widths by thickness effects rather
than duration effects.
  Explanations that depend on duration (e.g.,
growth of shocks), have a pulse width that depends on $\Delta t/\Gamma^2$
(see Table 1).  Thus, a constant peak width requires
that either $\Gamma$ changes in time as 
$\Delta t^{1/2}$ (no theory seems to predict this), or
 that $\Gamma$ and $\Delta t$
remain constant through the photon emission.
Although it is reasonable that $\Delta t$ is constant, most theories
expect $\Gamma$ to decrease as the shell loses energy.

Most of the above problems arise because the local spherical symmetry requires
weaker emission from material at $\theta \sim \Gamma^{-1}$ which will
arrive later by $\sim 0.2 T_0$.  The ``thick shell with substructure''
model (\S 3.4) is a promising way to overcome this and still have a single
release of energy at the central site.  If the shell thickens such that it is
much wider than the radius of curvature of the shell within $\Gamma^{-1}$,
it will act as a parallel slab, not a spherical surface.  A parallel
slab with embedded, small substructure that grows 
from a seed (as shocks might) could explain most GRBs time
history envelopes.

Table 2 summarizes various models and what sizes one obtains based on the
observed $T_D$ and $\Delta T_p$. The overall size of the region can
vary from $cT_D$ to $2\Gamma^2T_D$ and the perpendicular size of the emitting
region can vary from $c\Delta T_p$ to $2c\Gamma T_p$.

Merging neutron stars release their energy over a short period
of time and require an expanding shell to explain the
observed time variations.  The expansion can be relativistic to allow
the escape of photons well above the pair production threshold.
Our analysis indicates that for burst time histories, a central
engine explanation is preferred where the duration is dictated by the duration
of the energy release at the central site.
Failed supernova (Woolsey, 1993, Hartmann \& Woolsey 1995)
at cosmological distances might provide the central energy.
However, failed supernova do not produce the super relativistic shells
necessary to explain the high energy emission.
Alternatively, neutron stars in the halo of our Galaxy 
could provide the central site
since the energy release within any peak is not 
large enough to destroy the object.
The requisite relativistic expansion is smaller at distances commensurate
with halo models and can be produced with a much
smaller $\Gamma$ (see Nayakshin \& Fenimore 1996).

If GRBs are central engines, then each peak is probably due to a relativistic
expanding shell.  In that case, the results of this paper can be applied
to the individual peaks.  Most peaks have faster rises than falls
(Norris et al, 1996) and equation (\VTHICK) could be
used to estimate their characteristics.
The rise of the pulse is related to the duration of the photon-active phase
and the fall is related to the duration of the photon-quiet phase.

This work was done under the auspices of the US 
Department of Energy and was funded
in part by the {\it Gamma Ray Observatory} Guest Investigator Program.
We thank C.~Dermer, H.~Li, C.~Ho, P.~\Mesz for useful 
suggestions and corrections.

\centerline{\bf REFERENCES}

\BB
Epstein R. I., Fenimore E. E., Leonard P. J. T. \& Link B., 1993 Annals of
NY Acad. Sci., 688, 565

\BB
Fenimore E. E., et al. \apthree~158.

\BB
Fenimore E. E., Epstein R. I. \& Ho C., 1993a Astron. Astrophys. Suppl. Ser,
97, 59

\BB
Fenimore, et al., 1993b, Nature 366, 40.

\BB
Hartmann, D.~H., \& Woolsey, S.~E., 1995, Adv.~Space Res., 15, No.~5, 143

\BB
Hurley, K.,~ et. al. 1995, Nature 372, 652

\BB
Li, H., \& Fenimore,~E.~E~, 1996,
ApJ, Submitted

\BB
Katz J. I., 1994, ApJ, 422, 248

\BB
Koshut, T. M., et al.~1995 ApJ 452, 145

\BB
Krolik, J.~H, \& Pier, E.~A., 1991, ApJ 373, 227

\BB
Lamb D. Q., 1995, PASP 107, 1152

\BB
Meegan C. A., Fishman, G. J., Wilson, R. B., Paciesas, W. S., Brock,
M. N., Horack, J. M., Pendleton G. N., \& Kouveliotou, C., 1992,
Nature, 335, 143

\BB
\Mesz P., \& Rees, M.~J., 1992, MNRAS 258, 41p

\BB
\Mesz P., \& Rees M. J., 1993, ApJ, 405, 278

\BB
\Mesz P., \& Rees M. J., 1994 MNRAS 269, L41

\BB
Nayakshin, S., \& Fenimore, E.~E., 1995, in preparation for ApJ.

\BB
Norris, J.~P., et al., 1996 ApJ, 459, 393

\BB
\Pacz B., 1986, ApJ 308, L43

\BB
\Pacz B., 1990, ApJ 363, 218

\BB
\Pacz B., 1995, PASP 107, 1176

\BB
Piran, T. \aiptwo~495

\BB
Piran T. \& Shemi A., 1993, ApJ, 403, L67

\BB
Piran T., Shemi A. \& Narayan R., 1993 MNRAS 263, 861

\BB
Rees M. J., 1966 Nature 211, 468.

\BB
Rees M. J. \& \Mesz P., 1994, ApJ, 430, L93

\BB
Sari, R., \& Piran, T., 1995, ApJ 455, L143.

\BB
Schmidt W. K. H., 1978, Nature, 271, 525.

\BB
Shaviv N. \& Dar A., 1995 MNRAS 277, 287

\BB
Shemi A., 1994, MNRAS 269, 1112

\BB
Waxman, E. \& Piran T., 1994 ApJ, 433, L85

\BB
Woods E. \& Loeb. A., 1995 ApJ, 453, 583

\BB
Woolsey, S.~E., 1993, ApJ, 405, 273

\vfill\eject
\centerline{\bf FIGURE CAPTIONS}

{\bf Fig \FIGDIVERSITY.}
The diversity of GRB time histories.
({\it a})
Burst 1546 is a sharp single spike with a similar rise and fall time.
({\it b}) Burst 1885 is a FRED {fast rise, exponential decay).
A relativistic shell can appear to the observer as a FRED although the decay
phase is actually a power law rather than exponential.
({\it c}) Burst 2856 is a long complex burst.  Note that the spikes are
about the same size at the beginning as at the end.

{\bf Fig~\FIGELLP.} The surface that produces photons
from a relativistically expanding shell
as seen simultaneously by
a distant observer.
The observer is located at right at infinity. The shell 
originated at point O. The dotted lines at $r_1$, $r_2$ 
and $r_3$ represent hypothetical distances at which shells could produce
photons.   $\Gamma$ is the Lorentz gamma-factor of the expansion
and T is the photon arrival  time in the detector counted from the moment
of the beginning of the explosion.

{\bf Fig.~\FIGTHICKDUR.} The distinction between duration and thickness.
A shell with thickness $\Delta r_{\parallel}$ emits photons ``A'' and
``B.''  After time $\Delta t$, photons ``a'' and ``b'' are
emitted.  The thickness effects are determined by photons ``A''
and ``B'' whereas duration effects are determined by photons
``A'' and ``a.''
Duration effects are contracted by $\Gamma^2$ whereas thickness effects
are not.  Thus, the delay due to the curvature ($d_c$) is not contracted.

{\bf Fig \FIGVDELTA.}
Time history expected from a relativistic shell that generates gamma-rays
after expanding in a gamma-quiet phase for
time $T_0$.
({\it a}) The shape is a fast rise with a power law decay. 
 The width of the time structure
is proportional to the time spent in the gamma-quiet phase.
The dotted lines show a possible time history if only a single small
patch on the shell becomes active.
({\it b}) Relative boosting of the photons from the frame comoving
with the shell.  Photons at the peak of the time history originate
from the portion of the shell that is closest to the
observer.  Photons from the decay phase originate
off axis and the Lorentz boost is smaller by $T/T_0$.

{\bf Fig.~\FIGEMITSHELL.} The relationship between observed time structure
and the structure of the emitting region for the single, thin
shell model of the relativistic expansion.  In the top
panel, the elliptical curves are the surfaces that produce
gamma-rays that are seen at the same time at the detector.  Five gamma-ray
emitting regions produce peaks.
The width of the individual peaks is related to the size of the
individual emitting regions.
The overall duration of the event is related
to the overall size of the expansion.

{\bf Fig.~\FIGTHICKSUB.} The relationship between observed time structure
and the structure of the emitting region for
a shell which has grown to be very thick and develops substructure
within the shell.
If the thickness ($\Delta R_{\parallel}$) is larger than the radius of
curvature within $2\Gamma^2$ (i.e., $d_c$), then the time structure will
be dominate by the thickness and not local spherical symmetry.
The duration of the event is determined from the thickness of the
shell and the individual peaks are determined from the subregions that
develop into gamma-ray producing areas.

{\bf Fig.~\FIGFITEX.}
Fits of expected signals from relativistic shells to GRB time histories.
({\it a}) Burst 678 has a FRED-like envelope but many individual peaks.
The solid curve is the expected signal if the shell started to expand at
time 0.  The arrow points to the time when the spectrum should be
softer by a factor of 2.
The time structure should also be dilated by a factor of 2.
({\it b}) The dashed line is a fit of a relativistic shell to Burst 219.
The deduced start time for the shell is at time 0.
  This burst
had a precursor 100 s before this peak.  The solid line is the
expected shape assuming the shell started at the time of the precursor.
Either only a very small amount of the shell was converted into gamma-rays
or this is an example of a central engine, that is, multiple shells
spread out over 100 s.
({\it c}) Burst 2831, a long complex burst.
The dashed line is an expected shape from a single shell that becomes
gamma-active for a short time.  The dotted curve is the expected shape
for a shell that becomes active over a range of times.  The five solid
curves are the shapes expected if a single shell becomes active at five
different times.
In all cases, only a small fraction of the shell's surface generates the
gamma-rays.
This demonstrates the ``Shell Symmetry'' problem: the envelopes of most
GRB time histories
are not consistent with local spherical symmetry.
\vfill\eject
{}~~~~~~~
\baselineskip 14pt
\bigskip
\bigskip
\centerline{TABLE 1}
\centerline{Contributions to the Arrival Time of Photons from Duration
of}
\centerline{Emission and Thickness of Emitting Regions in Various Scenarios}
\def\tablerule{\noalign{\hrule}}
\vbox{\tabskip=0pt
\halign to 5.5truein{\strut#\tabskip=1em plus2em&
#\hfill&
#\hfill&
#\hfill&
#\hfill\cr
\tablerule
\cr
\tablerule

  & Patches  &  Seed Growth  &  Ambient Objects  \cr \cr
\tablerule
$\Delta T_{\Delta t}$  & $\Delta t/2\Gamma^2$ & $\Delta t/2\Gamma^2 $ &
$\Delta r_{\rm amb} /2 \Gamma^2 c$  \cr \cr
$\Delta T_{\Delta r_{\parallel}}$  & $\Delta r_{\parallel}/c$ & ${c_s'
\Delta t \over c \Gamma^2}$   & $ {{\rm min} (\Delta r_{\parallel}, \Delta
r_{\rm amb}) \over c}$ \cr \cr
$\Delta T_{\Delta r_{\perp}}$ & ${\Delta r_{\perp}^2 \over 4 \Gamma^2 c^2
T}$ & $ {c_s'^2 \Delta t^2 \over 4 \Gamma^6 c^2 (T+T_0)} $&
 $\DRamb^2/(4\Gamma^2 c^2 T)
$ \cr \cr
$\Delta T_p$  &
max $\biggl({\Delta r_{\parallel} \over c},{\Delta t \over 2
\Gamma^2},$&
 ${\Delta t \over \Gamma^2}((1/2)^2 +
(c_s'/c)^2)^{1/2}$
&$\Delta T_{\DRperp}$ or\dag~$\Delta T_{\DRpar}$
 \cr \cr
   & ${\Delta r_{\perp}^2 \over 4 \Gamma^2 c^2 T} \biggr)$
 & \cr \cr
\tablerule}}
\noindent
\dag~Depends on if object is collapsible or not, see text.
\hfill
\baselineskip 20pt

\medskip
\vfill\eject
{}~~~~~~~
\baselineskip 14pt
\bigskip
\bigskip
\centerline{TABLE 2}
\centerline{Size Estimates from Various Models\dag}
\def\tablerule{\noalign{\hrule}}
\vbox{\tabskip=0pt
\halign to 5.5truein{\strut#\tabskip=1em plus2em&
#\hfill&
#\hfill&
#\hfill\cr
\tablerule
\cr
\tablerule

   Model    &  $R$  &  $\Delta R_{\perp}$  \cr \cr
\tablerule
Thin Shell\hfill  &  $2c\Gamma^2 T_D $  &  $2c\Gamma T_D$  \cr \cr
Shell with Patches from\hfill  &
  $2c\Gamma^2 T_D$  &  $2c\Gamma(T_p \Delta T_p)^{1/2}$  \cr
simultaneous conditions\hfill && \cr  \cr
Shell with Patches from seeds\hfill&
  $2c\Gamma^2 T_D$  &  $c\Gamma \Delta T_p$  \cr \cr
Thick shell from\hfill &
 $2cT_D\big({ 1 \over \Gamma^2_{\rm min}}
-{1 \over \Gamma^2_{\rm max}}\big)^{-1}$    &
  $cT_b$,  $c\Gamma\Delta T_p$  \cr
spread in $\Gamma$\hfill && \cr \cr
Thick shell from\hfill & $>cT_b$    &  $cT_b$,  $c\Gamma\Delta T_p$  \cr
central engine\hfill && \cr \cr
Multiple Shells from \hfill  &
  $2c\Gamma^2 \Delta T_p$  &  $2c\Gamma \Delta T_p$  \cr
central engine\hfill \cr \cr
Stationary photosphere\hfill  &
  $c\Gamma \Delta T_p$   &  $c\Delta T_p$ \cr \cr
\tablerule}}
\noindent
\dag~$\Delta R_{\parallel}$ must always be $<c\Delta T_p$.
\hfill
\baselineskip 20pt

\medskip
\vfill\eject
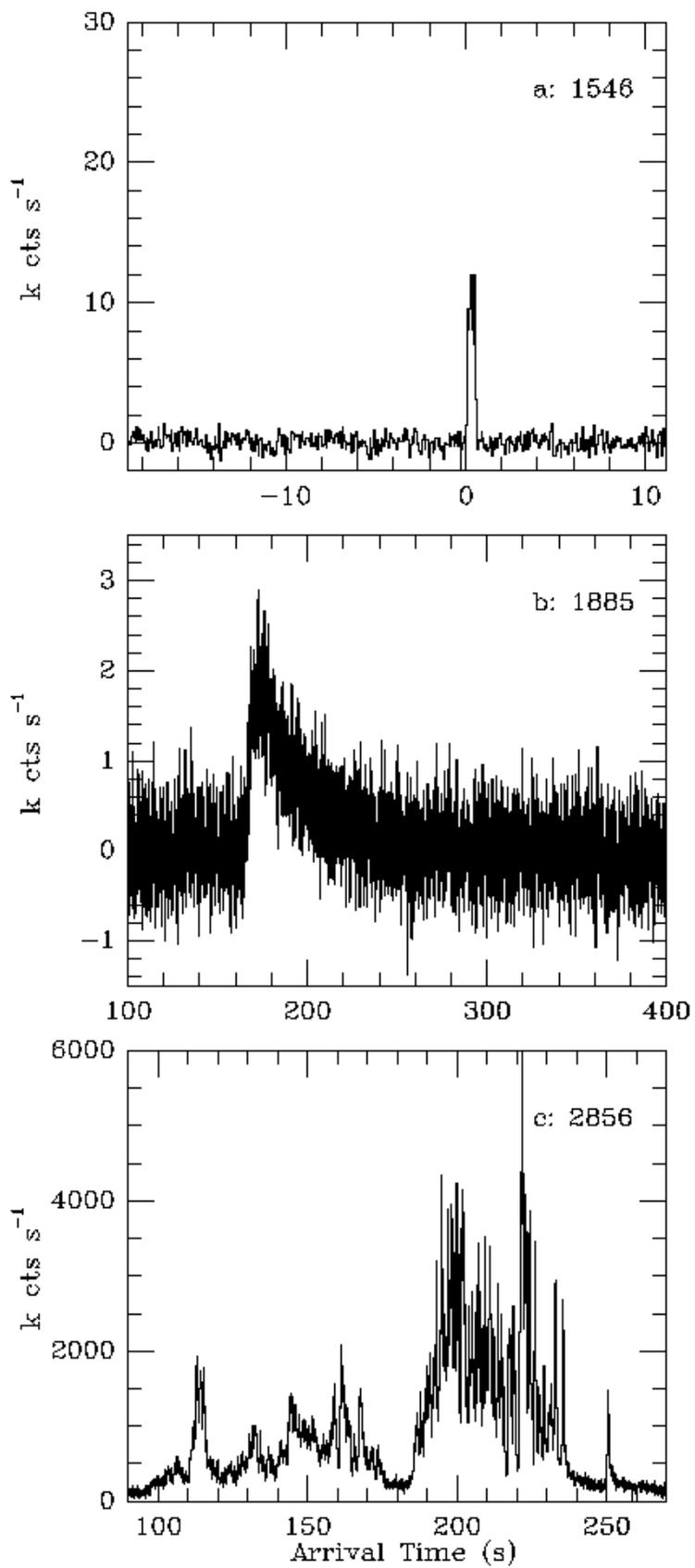
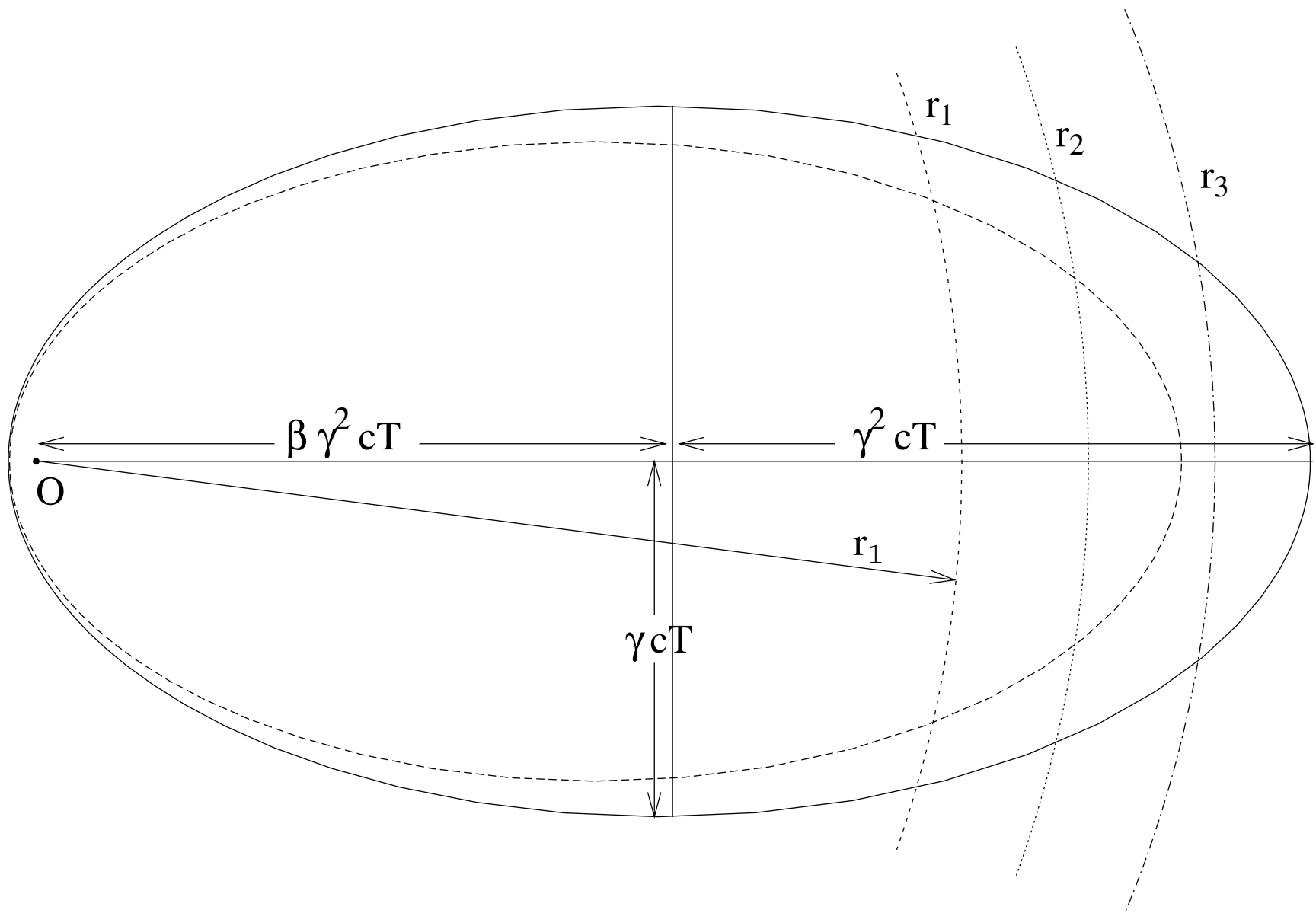
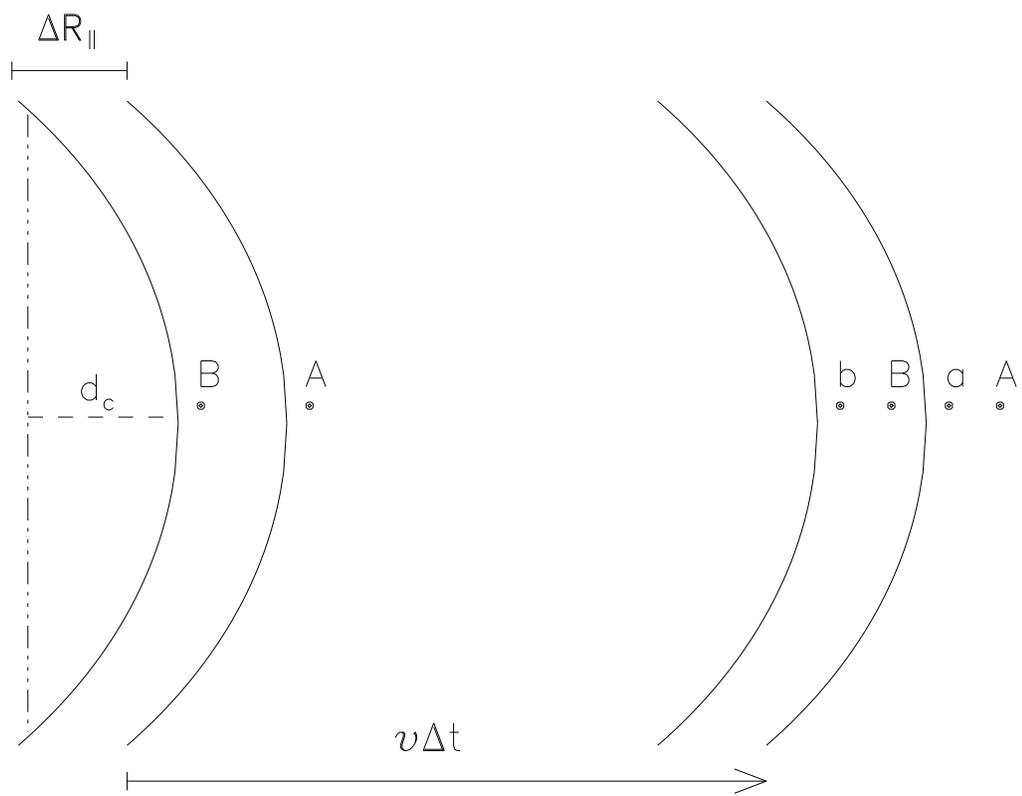
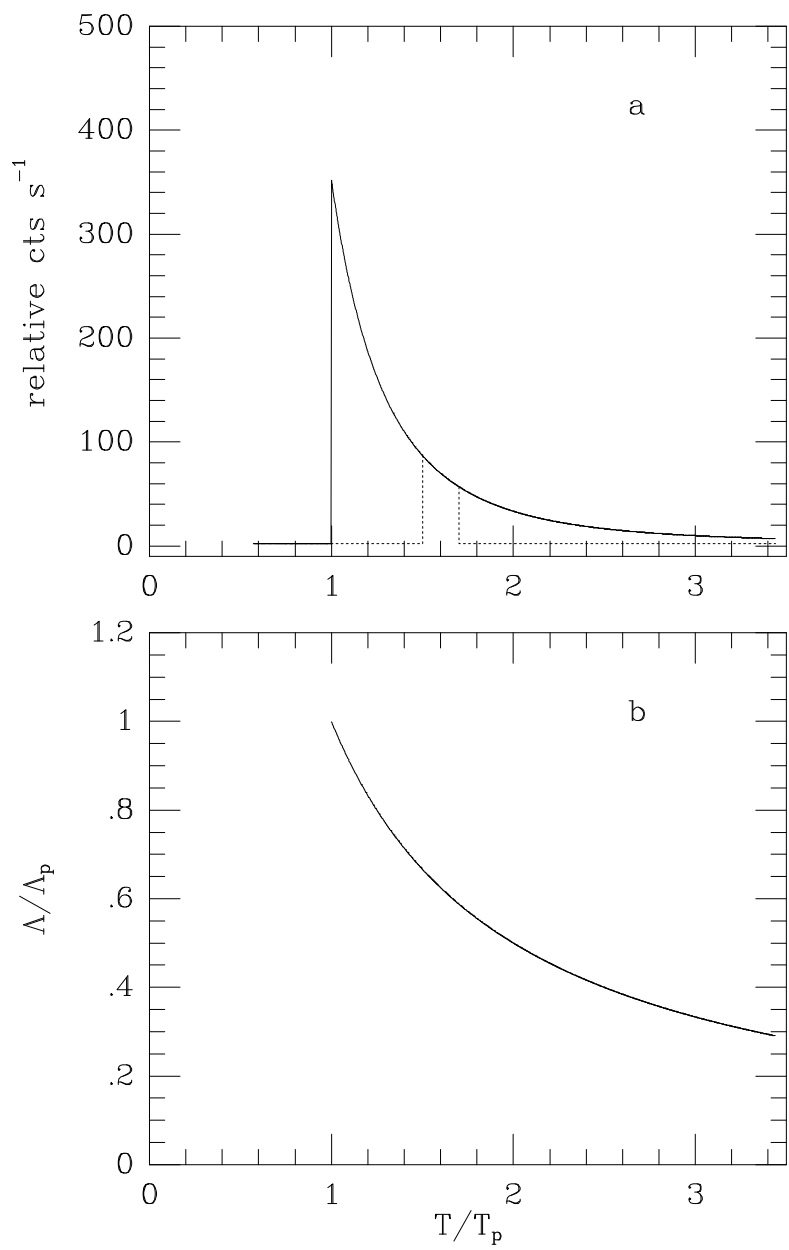
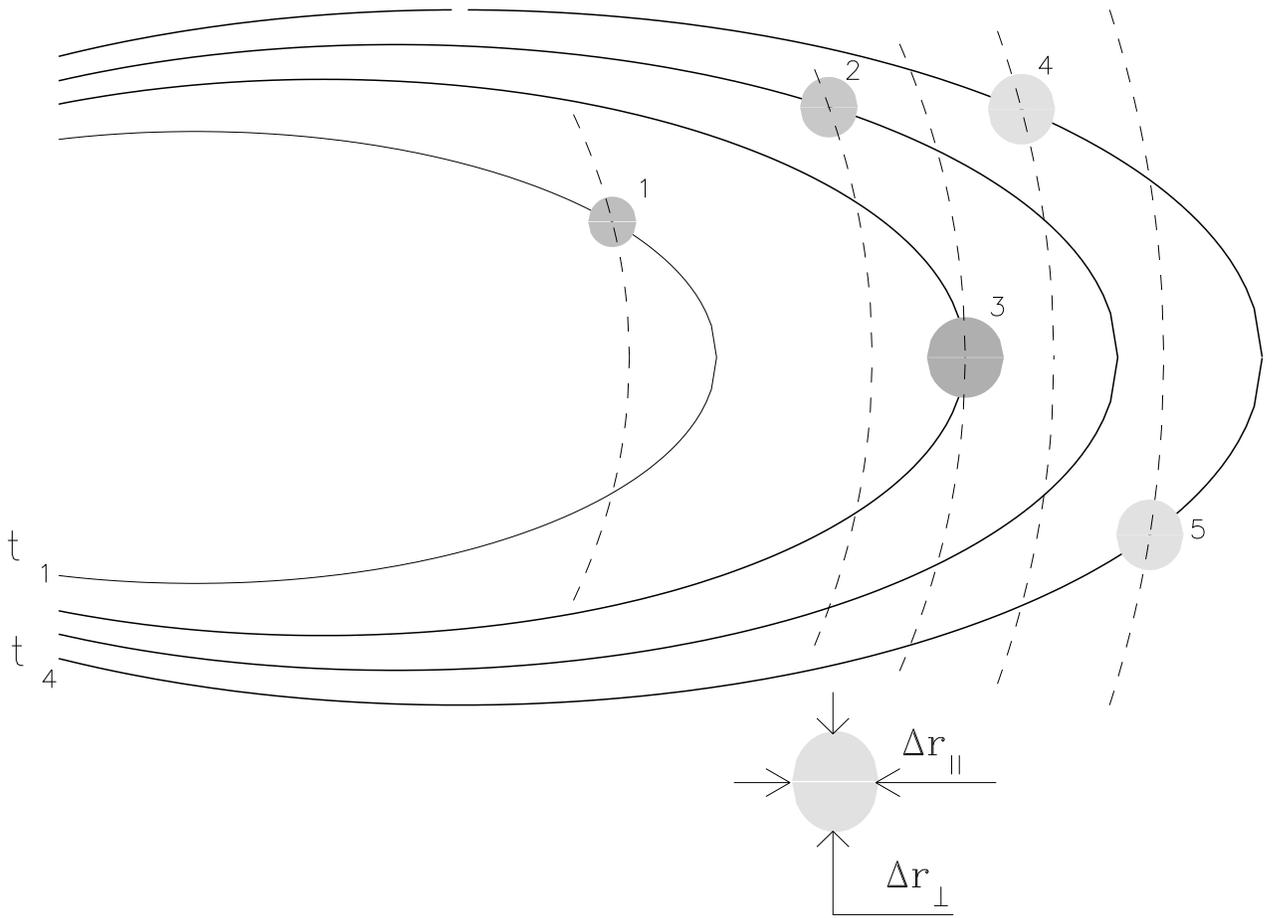
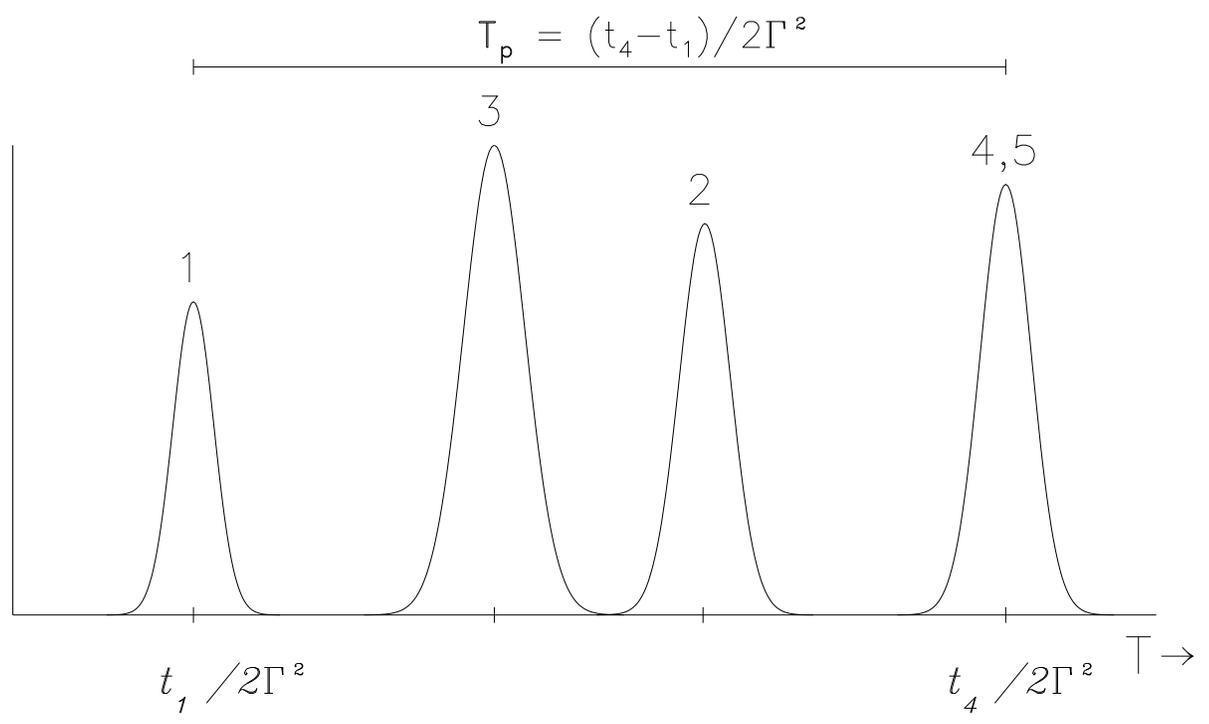
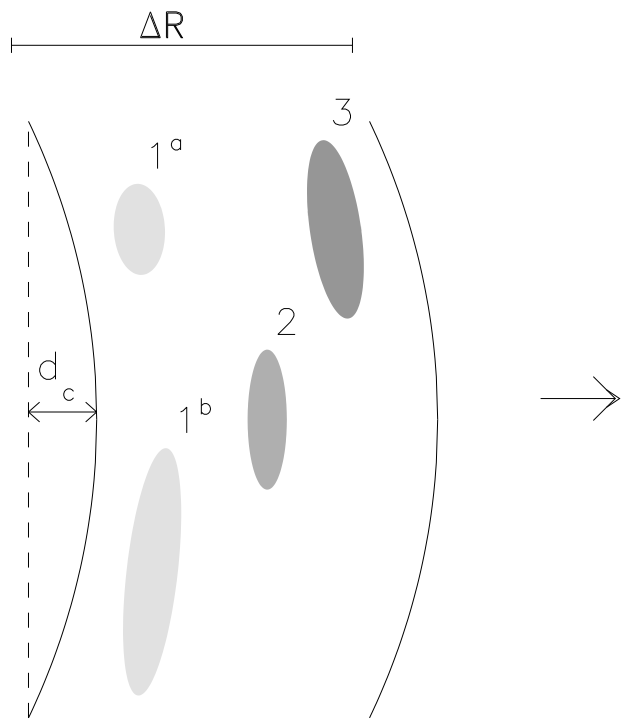
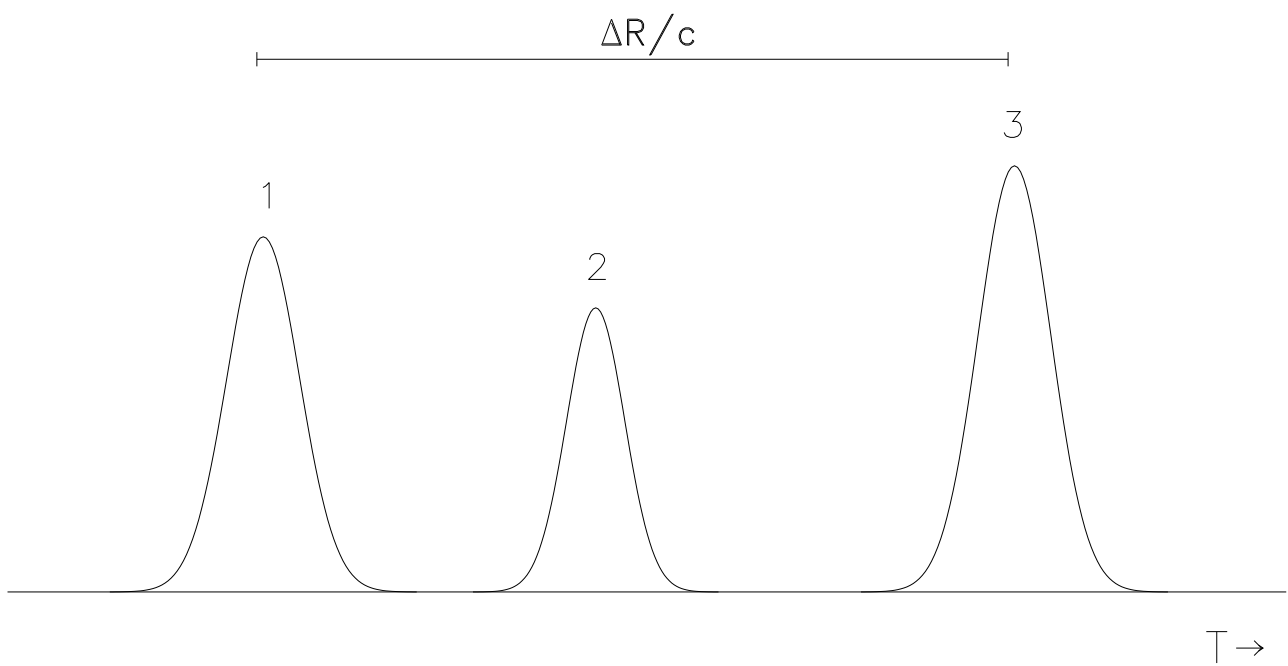
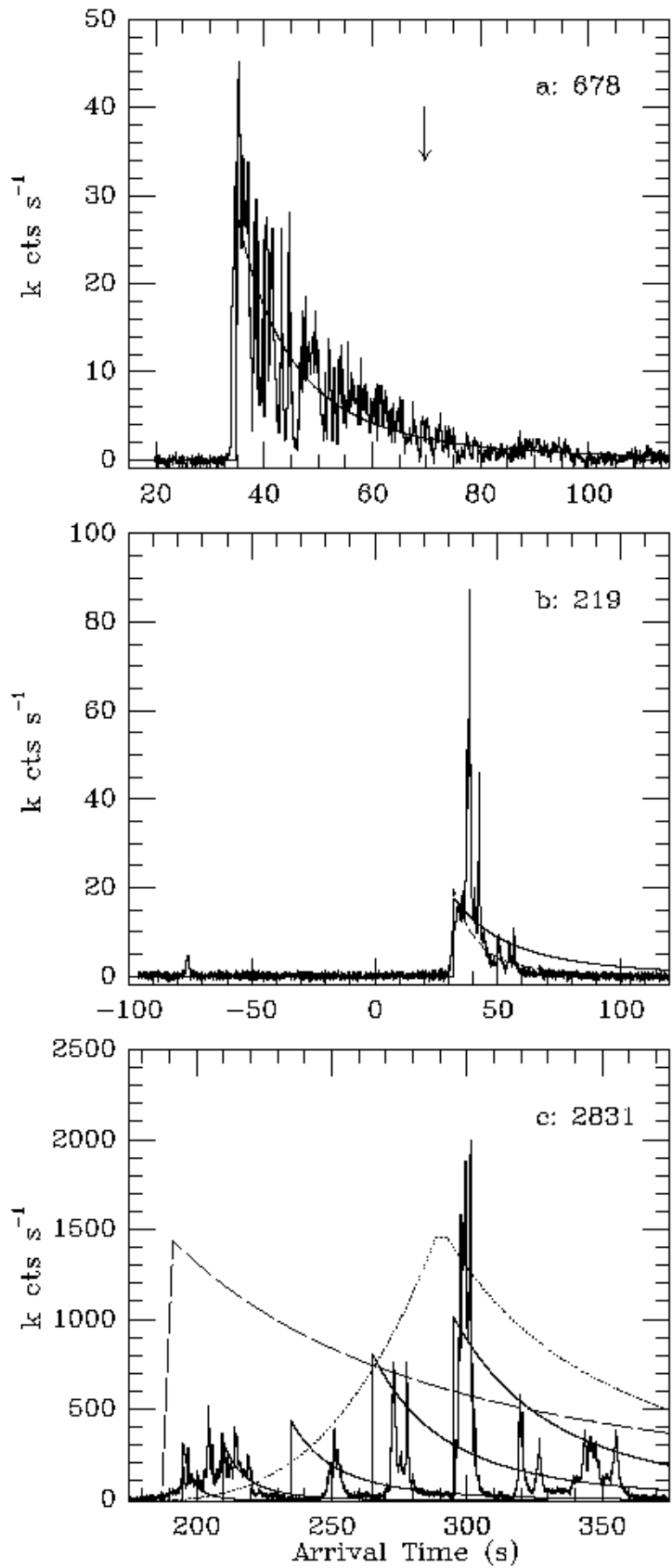
\bye